\documentclass[twocolumn,amsmath,trackchanges]{aastex702}
\usepackage{booktabs}
\usepackage{natbib}
\usepackage{amsmath}

\received{\today}
\revised{--}
\accepted{--}

\shorttitle{SN2024kgi}
\shortauthors{Dukiya et al.}

\newcommand{\ha}{H$\alpha$ }
\newcommand{\pa}{Pa$\alpha$ }
\newcommand{\hb}{H$\beta$ }

\newcommand{\power}[1]{$^{#1}$}

\usepackage{soul}
\newcommand\N[1]{{\color{blue} \bf #1}}

\newcommand{\ARIES}{\affiliation{Aryabhatta Research Institute of Observational Sciences, Manora Peak 263001, India}}
\newcommand{\MJPRU}{\affiliation{Department of Applied Physics, Mahatma Jyotiba Phule Rohilkhand University, Bareilly, 243006, India}}
\newcommand{\LCO}{\affiliation{Las Cumbres Observatory, 6740 Cortona Drive, Suite 102, Goleta, CA 93117-5575, USA}}
\newcommand{\UCSB}{\affiliation{Department of Physics, University of California, Santa Barbara, CA 93106-9530, USA}}
\newcommand{\hopkins}{\affiliation{Johns Hopkins University, San Martin Dr, Baltimore, MD 21210, USA}}
\newcommand{\UCD}{\affiliation{Department of Physics and Astronomy, University of California, Davis, 1 Shields Avenue, Davis, CA 95616-5270, USA}}
\newcommand{\seti}{\affiliation{SETI Institute, 189 Bernardo Avenue, Ste. 200, Mountain View, CA 94043, USA}}
\newcommand{\SNU}{\affiliation{Department of Physics and Astronomy, Seoul National University, Gwanak-ro 1, Gwanak-gu, Seoul 08826, Republic of Korea}}
\newcommand{\UA}{\affiliation{Steward Observatory, University of Arizona, 933 North Cherry Avenue, Tucson, AZ 85721-0065, USA}}
\newcommand{\GeminiNorth}{\affiliation{Gemini Observatory, 670 North A`ohoku Place, Hilo, HI 96720-2700, USA}}
\newcommand{\LSSTCatalyst}{\altaffiliation{LSSTC Catalyst Fellow}}
\newcommand{\SU}{\affiliation{Oskar Klein Centre, Department of Astronomy, Stockholm University, AlbaNova, SE-106 91 Stockholm, Sweden}}
\newcommand{\UTA}{\affiliation{Department of Astronomy, The University of Texas at Austin, 2515 Speedway, Stop C1400, Austin, TX 78712, USA}}
\newcommand{\INAF}{\affiliation{INAF - Osservatorio Astronomico di Padova, Vicolo dell’Osservatorio 5, 35122 Padova, Italy}}
\newcommand{\INAFBrera}{\affiliation{INAF - Osservatorio Astronomico di Brera, Via Bianchi 46, 23807 Merate (LC), Italy}}

\begin{document}

\title{Early NIR echo and the time variable mass loss history of Type IIn SN 2024kgi}

\correspondingauthor{Kuntal Misra}
\email{kuntal@aries.res.in}

\author[0000-0002-0394-6745]{Naveen Dukiya} \ARIES \MJPRU
\email{ndookia@gmail.com}

\author[0000-0003-1637-267X]{Kuntal Misra} \ARIES
\email{kuntal@aries.res.in}

\author[0000-0002-7259-4624]{Andrea Pastorello} \INAF
\email{andrea.pastorello@inaf.it}

\author[0000-0003-4254-2724]{Andrea Reguitti} \INAF \INAFBrera
\email{andrea.reguitti@inaf.it}

\author[0000-0002-3884-5637]{Anjasha Gangopadhyay} \SU
\email{anjashagangopadhyay@gmail.com}

\author[orcid=0000-0002-7352-7845, gname=Aravind, sname=Ravi]{Aravind P.\ Ravi}
\UCD \email{apazhayathravi@ucdavis.edu}

\author[orcid=0000-0002-9454-1742, gname=Brian, sname=Hsu]{Brian Hsu} \UA
\email{bhsu@arizona.edu}

\author[0000-0003-4537-3575]{Noah Franz}
\UA \email{nfranz@arizona.edu}

\author[0000-0003-4102-380X]{David J. Sand}
\UA \email{dsand@arizona.edu}

\author[orcid=0000-0001-5510-2424, gname=Nathan, sname=Smith]{Nathan Smith}
\UA \email{nathansmith@arizona.edu}

\author[0000-0003-0123-0062]{Jennifer E. Andrews}
\GeminiNorth \email{Jennifer.Andrews@noirlab.edu}

\author[0000-0002-1895-6639]{Moira Andrews} \LCO 
\email[]{mandrews@lco.global}

\author[0000-0002-4924-444X]{K. Azalee Bostroem} \UA \LSSTCatalyst
\email[]{bostroem@arizona.edu}

\author[0000-0003-0528-202X]{Collin Christy}
\UA \email{collinchristy@arizona.edu}

\author[0009-0002-2621-6611]{Monalisa Dubey} \ARIES \MJPRU
\email{monalisa@aries.res.in}

\author[0000-0003-4914-5625]{Joseph R. Farah} \LCO \UCSB
\email[]{jfarah@lco.global}

\author{Archana Gupta} \MJPRU
\email{}

\author[0000-0003-0209-9246]{Estefania Padilla Gonzalez} \hopkins
\email[]{epadill7@jh.edu}

\author[orcid=0000-0003-2744-4755, gname=Emily, sname=Hoang]{Emily Hoang} \UCD \email{emthoang@ucdavis.edu}

\author[0000-0003-4253-656X]{D. Andrew Howell} \LCO \UCSB
\email{dahowell@gmail.com}

\author[0000-0001-5807-7893]{Curtis McCully} \LCO \UCSB
\email{cmccully@lco.global}

\author[orcid=0009-0008-9693-4348, gname=Darshana, sname=Mehta]{Darshana Mehta}
\UCD \email{ddmehta@ucdavis.edu}

\author[0000-0001-9570-0584]{Megan Newsome} \UTA
\email[]{newsome.megane@gmail.com}

\author[0000-0002-0744-0047]{Jeniveve Pearson}
\UA \email{jenivevepearson@arizona.edu}

\author[orcid=0000-0002-7015-3446, gname=Nicol\'as, sname=Meza Retamal]{Nicol\'as Meza Retamal}
\UCD \email{nemezare@ucdavis.edu}

\author[0000-0003-3643-839X]{Jeonghee Rho} \seti \SNU
\email{jrho@seti.org}

\author[0000-0002-4022-1874]{Manisha Shrestha}
\UA \email{manisha.shrestha@monash.edu}

\author[0000-0001-8073-8731]{Bhagya Subrayan}
\UA \email{bsubrayan@arizona.edu}

\author[0000-0003-0794-5982]{Giacomo Terreran}
\affiliation{Adler Planetarium, 1300 S DuSable Lake Shore Dr, Chicago, IL 60605, USA} \email[]{gterreran@adlerplanetarium.org}

\author[0000-0002-3697-2616]{Lina Tomasella} \INAF
\email{lina.tomasella@inaf.it}

\author[orcid=0000-0001-8818-0795, gname=Stefano, sname=Valenti]{Stefano Valenti}
\UCD \email{valenti@ucdavis.edu}

\author[0000-0002-3334-4585]{Giorgio Valerin} \INAF
\email{giorgio.valerin@inaf.it}

\author[0009-0006-7296-728X]{Kathryn Wynn} \LCO \UCSB
\email{kwynn@lco.global}

\begin{abstract}
We present a comprehensive analysis of the long-term photometric and spectroscopic monitoring campaign of the Type IIn Supernova (SN) 2024kgi. The SN reaches at peak an $r$-band absolute magnitude of $-19.81 \pm 0.06$ mag and a bolometric luminosity of $\sim 2 \times 10^{43}$ erg s$^{-1}$. 
We observe a break in the lightcurve at day $\sim 312$, after which the luminosity decline changes from $t^{-1.0}$ to $t^{-4.5}$, likely due to the shock sweeping the dense CSM.
We introduce a semi-analytical lightcurve modeling approach for CSM-interaction in transients that accounts for variable diffusion time. 
We hence estimate a CSM mass of $2.34^{+11.8}_{-1.6} M_{\odot}$ and with a density profile $\rho_{csm} \propto r^{-2.74}$, indicating progressively increasing mass loss toward the SN explosion. 
The ejecta signatures emerge as broad H, He, and \ion{Ca}{2} triplet lines, at day $\sim 80$, much earlier than expected from a spherically symmetric CSM, suggesting asymmetry in the CSM. 
After the lightcurve break, the H lines exhibit a wavelength-dependent deficit in the red-wing flux, indicating new dust formation in the ejecta and/or in the post-shock gas. We observe an NIR excess from day $\sim 40$ onwards. We attribute the early NIR excess before the lightcurve break solely to the NIR echo from pre-existing dust, as there are no other indications for new dust formation. The late-time NIR excess likely has contributions from both the preexisting and newly formed dust. The steady increase in mass loss, slow CSM velocity, and asymmetric CSM favor a binary interaction-induced mass loss.
\end{abstract}




\section{Introduction} \label{sec:intro}
Massive stars ($>8\,M_{\odot}$) end their life as core-collapse supernovae (SNe; \citealp{woosley_heger_massive_star_review_2002}). A subset of these SNe, showing narrow Balmer lines throughout most of their evolution, are known as Type IIn SNe (SNe~IIn hereafter; \citealp{schlegel1990_IIn, filippenko97_review, smith_IIn_review, fraser_interacting_transients_2020}). The narrow lines originate from a dense photoionized circumstellar medium \citep{chugai_danziger_1994}.
The H lines in the early spectra typically show Lorentzian wings in addition to a narrow feature, characteristic of electron-scattering in the unshocked CSM. Later in the evolution, the lines can develop a multi-component profile, showing narrow, intermediate, and broad component, likely originating from the unshocked CSM, the post-shock cool dense shell (CDS), and the SN ejecta, respectively. These components can be observed simultaneously or at different times during evolution, depending on the CSM and the explosion geometry \citep{chugai_danziger_1994, Dessart_IIn_simulation,smith_IIn_review}.
Additionally, the H lines can exhibit asymmetries during evolution due to CSM/explosion asymmetry, obscuration by the continuum photosphere, or newly formed dust in the ejecta/CDS \citep{2012ab_gangopadhyay, 2013L_taddia, 2017hcc_smith_andrews, 2015da_smith}. 

Another important aspect of SNe~IIn is the presence of dust in the pre-existing CSM (which gives rise to dust-echoes; \citealp{Fox_spitzer_IIn_2011, Dwek_dust_echo_2010jl_2021}), and the dust formation in the post-shock gas and/or the ejecta (which gives rise to late time NIR excess and suppression of red-wing in H line profiles; \citealp{pozzo_1998s_dust, 2010jl_Sarangi_2018, 2017hcc_smith_andrews, 2015da_smith}). The formation of a CDS enables highly efficient radiative cooling and facilitates the rapid condensation of large amounts of new dust \citep{Smith_2006jc_dust_2008, Smith_2005ip_dust}, which may alleviate the dust budget tension in early-type galaxies \citep{dust_budget_rowlands_2014, sarangi_dust_formation_cds_2022}.

The CSM is usually generated by the progenitor star through mass loss in the years to decades prior to the explosion \citep{Smith_mass_loss_2014}. Sometimes eruptive mass loss can occur right before a SN explosion \citep{smith10,mauerhan13,2009ip_pastorello, pastorello_2016bdu, pastorello_2018cnf, 2023ldh_pastorello}, often referred to as precursor activity. \cite{reguitti_precusor_2024} find precursor activity in 29\% of nearby SNe IIn in their sample, consistent with the findings of \cite{strotjohann_precursor_rates_2021}, who reported that 26\% of SNe IIn exhibit precursors brighter than $-13$ mag in the $r$ band within the final three months before explosion.
Studying these events can reveal the progenitor star's mass loss history and place constraints on the physical mechanisms responsible for it. 
Some of the proposed mechanisms for mass loss include very strong line-driven winds \citep{1995N_fransson, smith_2009_vy_canis_majoris}, wave-driven oscillations close to the core-collapse \citep{Quataert_wave_driven_massloss_2012, Shiode_wave_driven_massloss_2014, Wu_fuller_wavedriven_mass_loss_2021}, pulsational pair instabilities \citep{Woosley_ppisn_2017, Woosley_smith_1961V_2022}, nuclear burning instabilities \citep{arnett_nuclear_instabilities_2011,sa14}, and binary interactions \citep{sa14,Smith_eta_carinae_light_echo_2018, Schorder_binary_IIn_2020}. 
These mechanisms predict different timescales, CSM geometry and density profiles, and wind velocities; estimating these parameters may help constrain the mass-loss mechanisms.

SNe~IIn are intrinsically rare, constituting $\sim 9\%$ of all core-collapse SNe \citep{smith11frac,Li_lick_sn_rates}. They form a heterogeneous and diverse class of transients. \cite{hiramatsu_IIn_optical} study the optical multiband lightcurves of over 450 SNe~IIn and find a median $r$ band absolute magnitude of $-19.16^{+1.33}_{-1.34}$ mag (similar to a study by \citealp{nyholm_IIn_survey_2020}). They also find a bimodality in the luminosity-timescale phase-space -- a luminous-slow group, and a faint-fast group with median total radiated optical luminosities $\sim 2 \times 10^{50}$ erg, and $\sim 10^{49}$ erg, respectively. This could indicate distinct progenitor channels for these SNe. Many authors have developed analytical and numerical modeling frameworks to explain the explosion physics and observables of SNe~IIn \citep{smith10gy,CW12, takashi_IIn_analytical, Dessart_IIn_simulation, Tsuna_IIn_model_2019, Suzuki_2D_CSM_2019, Khatami_Kasen_2024, ransome_villar_mosfit_sample_IIn, Transfit-CSM_zhang_2026, Sarin_redback_interacrting}. The modeling approaches indicate massive CSM (up to tens of solar masses) and high energy budgets for luminous-slow SNe~IIn, whereas the faint-fast group can be explained by interaction with a relatively modest amount of CSM.  \cite{hiramatsu_IIn_optical} speculate that the possible progenitor for the faint-fast group can be Red Super Giants and Asymptotic Giant Branch stars \citep{Smith_RSG_IIn_2009, Moriya_RSG_IIn_2011, Burrows_IIn_2024},
while the slow-luminous group best matches expectations for terminal explosions of Luminous Blue Variables \citep{Humphreys_davidson_LBV_1994, Smith_mass_loss_2014,smith26}, 
massive interacting binaries \citep{Chevalier_binary_IIn_2012,sa14,Schorder_binary_IIn_2020}, or even pulsational-pair instability driven SNe \citep{Woosley_ppisn_2017, Woosley_smith_1961V_2022}.

While observations over the past decades have established CSM interaction as the defining characteristic of SNe IIn, the physical properties and formation mechanisms of their dense circumstellar environments remain poorly constrained.
In this context, we present a case study of SN~2024kgi, with long-term photometric and spectroscopic coverage, allowing us to investigate these aspects in detail.
The paper is structured as follows. The observations and data reduction procedures, along with key SN parameters (explosion epoch, extinction, and distance), are described in \autoref{sec:observations}. The optical lightcurve is presented in \autoref{sec:lightcurve}, and in \autoref{sec:bolometric}, we analyze the properties and provide a model of the bolometric lightcurve. \autoref{sec:spectral_evolution} describes the overall spectral properties and the detailed evolution of the H line profiles. We discuss the key implications of our analysis in \autoref{sec:discussion}, and conclude the results in \autoref{sec:conclusions}.
 
\section{Observations and Data Reduction} 
\label{sec:observations}
SN~2024kgi was discovered by the Gravitational-wave Optical Transient Observer (GOTO) at RA = 22:44:43.985, DEC = +16:05:06.95 on UT 2024-06-03 05:05:47.904 (JD 2460464.71) at a magnitude of $18.61 \pm 0.10$ (ABmag) in the GOTO-L band \citep{discovery_report}. The GOTO team reports a non-detection in the GOTO-L band on JD 2460460.71 with a limiting magnitude of 19.4 ABmag. 
The transient was classified by \cite{classification_report} as a Type IIn SN at a redshift of $\sim 0.04$, using a spectrum taken with the SNIFS instrument mounted on the University of Hawaii 88-inch telescope. The SN is located at the outskirts of WISEA J224443.37+160504.5, its potential host galaxy.

\subsection{Photometric Observation}
SN~2024kgi was observed in the $UBVgri$ bands with the Las Cumbres Observatory (LCO) network of telescopes as part of the Global Supernova Project (GSP). The 1-m class LCO telescopes were used for imaging observations. All these telescopes offer identical imaging systems, using 4k $\times$ 4k CCD detectors that cover a field-of-view of $15'.8 \times 15'.8$ \citep{LCO_telescopes_instruments}. The observations started from $\sim 11$ days and continued up to $\sim 570$ days after discovery. 
The pre-processing (bias correction, flat correction, cosmic-ray correction, and astrometry of the frames) was carried out using the \texttt{BANZAI} pipeline \citep{Banzai}. The photometry and calibration was performed using the \texttt{IRAF} based \texttt{lcogtsnpipe}\footnote{\url{https://github.com/LCOGT/lcogtsnpipe/}} pipeline \citep{valenti_lcogtsnpipe}. This pipeline calculates the instrumental magnitudes using the point-spread function (PSF) technique. The $UBV$ magnitudes of the local stars in the frame were derived by calibrating against Landolt fields \citep{landolt_2009} observed on the same night at several epochs. The $UBV$ instrumental magnitudes were then calibrated against these local stars for all frames. The $gri$ magnitudes were calibrated against the Sloan Digital Sky Survey (SDSS; \citealp{sdssdr17_abdurrouf}).

We also observed SN~2024kgi with the Asiago Faint Objects Spectrograph and Camera (AFOSC) on the Copernico 1.82m telescope in $gri$ bands, and with the Asiago Schmidt 67/92 telescope in $BVgri$ bands. PSF photometry was performed using a custom Python photometry pipeline (Dukiya et al., in preparation). The PSFs were derived using \texttt{PSFEx} \citep{psfex_ascl}, and the instrumental magnitudes were derived by fitting the detected sources with the PSF model using \texttt{photutils} \citep{larry_bradley_2024_photutils}. The $BV$ instrumental magnitudes were calibrated against the local sequence derived from the LCO data, and the {\em gri} instrumental magnitudes were calibrated against SDSS catalog stars \citep{sdssdr17_abdurrouf}. 

The Asteroid Terrestrial-impact Last Alert System (ATLAS; \citealp{atlas_tonry_2018}) regularly monitored the SN field covering epochs leading to the explosion. Forced photometry was performed on the location of the SN using the ATLAS forced photometry server\footnote{\url{https://fallingstar-data.com/forcedphot/}} \citep{atlas_forcedphot} to construct the SN lightcurve in the ATLAS $cyan$ (hereafter, $c$) and $orange$ ($o$) bands. We performed sigma-clipping and stacked the lightcurve data in 1-day bins using the method described by \cite{Young_plot_atlas_fp}. 

The photometric data are provided in \autoref{tab:photometry_log} and the complete lightcurve of SN~2024kgi is in \autoref{fig:lightcurve}.

\subsection{Spectroscopic Observations}

Low-resolution (R $\sim$ 400-700) optical spectroscopic observations were carried out with the cross-dispersed FLOYDS spectrographs mounted on the LCO 2m telescopes as a part of the GSP. The observations were taken at regular cadence to cover the spectroscopic evolution of SN~2024kgi. The \texttt{floydsspec}\footnote{\url{https://github.com/LCOGT/floyds_pipeline}} pipeline \citep{Valenti_floyds} was used to extract the 1D spectrum, to calibrate the wavelength with respect to a HgAr lamp spectrum taken after the exposure, and to calibrate in flux using a standard star observed on the same night. 

Low to medium resolution optical spectra were taken using the ARIES-Devasthal Faint Object Spectrograph and Camera (ADFOSC; \citealp{adfosc_omar, adfosc_dimple}) on 3.6m Devasthal Optical Telescope (DOT; \citealp{DOT_brijesh_2018}), the Himalayan Faint Object Spectrograph and Camera (HFOSC) on 2.0m Himalayan Chandra Telescope (HCT), the Calar Alto Faint Object Spectrograph (CAFOS) on the 2.2m telescope at Calar Alto observatory, the Asiago Faint Object Spectrograph and Camera (AFOSC) on the Copernico 1.82m telescope, and the Device Optimized for LOw RESolution (DOLORES) on Telescopio Nazionale Galileo (TNG). Two high-resolution spectra were taken using the Blue Channel spectrograph (BCH) on the 6.5m MMT telescope. The image pre-processing (bias correction, flat correction) for ADFOSC and HFOSC was done using standard procedures in \texttt{ccdproc}\footnote{\url{https://ccdproc.readthedocs.io/}} \citep{ccdproc_matt_craig_2017} \texttt{Python} library, and cosmic-ray removal was done using \texttt{astroscrappy} \citep{astroscrappy}. For the remaining instruments, the pre-processing and cosmic-ray removal were performed using standard tasks in \texttt{IRAF}. For all instruments, extraction of the one-dimensional (1D) spectrum, and wavelength calibration were performed using standard tasks in \texttt{IRAF}. The flux calibration was achieved using spectrophotometric standards observed at an airmass similar to that of each science frame. In the case of multiple exposures, the resulting spectra were median-combined into a single 1D spectrum for each epoch.

NIR spectra of SN~2024kgi were acquired using the SpeX spectrograph \citep{irtf_spex_rayner} on the NASA InfraRed Facility Telescope (IRTF). Data reduction of SpeX data was performed with \texttt{Spextool} \citep{irtf_spextool_cushing}. We obtained one epoch of NIR spectra ($zJ$) using the Magellan Infrared Spectrograph (MMIRS; \citealp{mmirs_mcleod}) on the 6.5m MMT telescope. The data were reduced using the MMIRS pipeline \citep{mmirs_pipeline_chilingarian}. Telluric bands correction and flux calibration were performed with \texttt{xtellcor} \citep{xtelcorr_vacca} using a standard A0V star observed at a similar time and a similar airmass.

The Spectro-Photometer for the History of the Universe, Epoch of Reionization and Ices Explorer \citep[SPHEREx;][]{spherex_satellite_bock} is a space telescope launched on 2025 March 11 to conduct an all-sky survey in 102 infrared colors between 0.75 and 5 $\mu$m, hence covering a wide range of wavelengths from the NIR to the mid-infrared (MIR) domains.
During the first two scannings of the whole celestial sphere, between MJD 60828 and 60842 ($\sim 369$~d\, to $\sim +383$~d), and again between MJD 60990 and 61020 ($\sim +531$~d to $\sim +561$~d), it observed the field of SN 2024kgi.
We retrieved the public calibrated spectrophotometry \citep{spherex_specphot_pipeline, spherex_spectral_response} from the NASA/IPAC Infrared Science Archive web page.\footnote{\url{https://irsa.ipac.caltech.edu/applications/spherex/tool-spectrophotometry}} 

We also include the publicly available classification spectra submitted to TNS \footnote{\url{https://www.wis-tns.org/object/2024kgi}} in our analysis \citep{classification_report}. The log of spectroscopic observation is provided in \autoref{tab:spectra_log}.

\subsection{Explosion Epoch}
\label{sec:explosion_epoch}
The ATLAS $o$-band lightcurve has the best cadence among all the bands and covers the rise of the SN. The earliest detection of the SN is on JD 2460463.08. We find a non-detection on JD 2460456.11 in the ATLAS $o$-band data at a limiting magnitude of 20.3 mag, but a GOTO non-detection on JD~2460460.71 provides a much tighter constraint on the rise. To estimate the explosion epoch of SN~2024kgi, we fit a power-law to the rising part of the flux observed in the ATLAS $o$ band. We perform the fitting using Markov-chain Monte Carlo (MCMC) simulations and find that the lightcurve rises as $f \propto (t - 2460459.8^{+1.1}_{-1.5})^{(1.20^{+0.2}_{-1.5})}$, where $t$ is the JD time. This rise is consistent with the non-detection on JD~2460460.71 reported by GOTO. From this, we estimate the explosion time to be JD $2460459.8^{+1.1}_{-1.5}$, and use it as a reference epoch throughout this paper.

\subsection{Extinction} 
\label{sec:extinction}
The Milky Way extinction along the line-of-sight to SN~2024kgi is E(B-V)$_{\mathrm{MW}} = 0.073$ mag \citep{milkyway_reddening}, as reported in the \cite{NED_extinction_calc}1\footnote{\url{https://ned.ipac.caltech.edu/byname?objname=SN+2024kgi}}. The spectra do not show any significant Na ID lines at the SN redshift, which is not unexpected as the SN lies in the outskirts of the host galaxy. Hence, the host galaxy extinction is negligible.

\subsection{Redshift and Distance} 
\label{sec:distance}
SN~2024kgi was reported with a redshift of $\sim 0.04$ \citep{classification_report}, but the redshift of the possible host galaxy is not available in the literature. However, from the narrow component of the \ha line, we can estimate the SN redshift with a fair precision. The narrow lines indicate a SN redshift of $0.0385 \pm 0.0001$ throughout the spectral evolution. 
Using the \texttt{cosmology} module of \texttt{Astropy}, we find that this redshift corresponds to a distance of $162.9 \pm 0.4$ Mpc ($\mu =  36.059 \pm 0.005$ mag;  assuming H$_0 = 73$ km/s/Mpc, $\Omega_{\mathrm{matter}}=0.27$, and $\Omega_{\mathrm{vaccum}}=0.73$), where the error only represents the error in redshift propagated to distance.

\section{Lightcurve Properties of SN2024kgi} \label{sec:lightcurve}

\begin{figure}
    \centering
    \includegraphics[width=\linewidth]{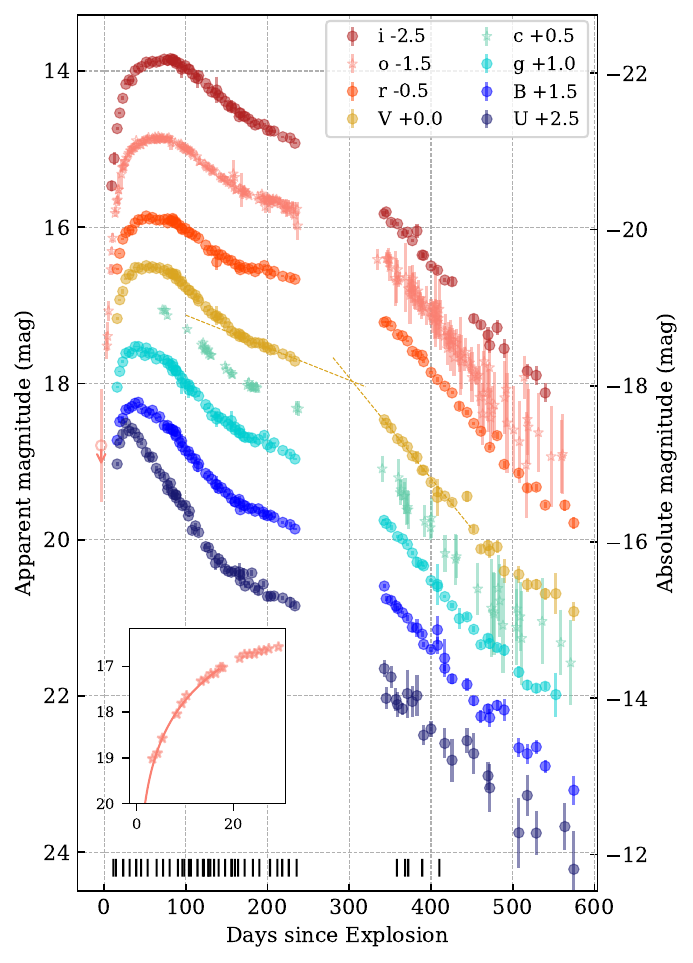}
    \caption{Multiband lightcurve of SN~2024kgi. Offsets have been applied to the lightcurves for clarity, as specified in the legend. The vertical ticks at the bottom represent the epochs when optical spectra are taken. The dashed line shows the straight line fits to $V$ band lightcurve used to estimate the decline rates (see \autoref{tab:peak_rise_slope}). The \textit{inset} shows the power-law fit to the $o$ band as discussed in \autoref{sec:explosion_epoch}}
    \label{fig:lightcurve}
\end{figure}

The $UBgcVroi$ lightcurve of SN~2024kgi is shown in \autoref{fig:lightcurve}. SN~2024kgi shows a mostly smooth photometric evolution.
We performed spline fitting in each band lightcurve to estimate the rise times and the peak magnitudes. Quartic smoothing splines were fitted with \texttt{Scipy} \citep{2020SciPy-NMeth}, with a smoothing factor chosen to avoid overfitting. The uncertainties associated with the fits and, in turn, with the rise times and peak magnitudes were determined using Monte Carlo experiments. We performed 1000 iterations of spline fittings, in which each data point varied within its photometric uncertainties. The rise time ranges from 41 days (for the $U$ band) to 75 days (for the $i$ band). The rise times, peak magnitudes, and decline rates at distinct phases of the SN evolution are provided for all bands in \autoref{tab:peak_rise_slope}. The peak absolute magnitudes are $< -19.5$ mag in all filters, slightly brighter than the median value for SNe~IIn \citep{hiramatsu_IIn_optical}. The lightcurves show a smooth post-maximum decline across all bands. The decline in the blue bands is faster than in the redder bands, with $U$ and $i$ bands declining at rates of $1.57 \pm 0.09$ mag/(100 d) and $0.93 \pm 0.03$ mag/(100 d), respectively. This behavior is expected as the expanding photosphere cools down. The $r$ and $o$ bands are exceptions to this trend, and decline much more slowly compared to all other bands, likely due to the presence of a very strong H$\alpha$ emission (see \autoref{fig:bolometric}).

After day 80, the magnitudes decrease linearly with the log of time, characteristic of ongoing CSM interaction \citep{Khatami_Kasen_2024}. In \autoref{fig:lightcurve}, this is seen as a flattening in all bands after day $\sim 160$ relative to the initial decline. The decline during this flattening phase is much more consistent among the optical bands, at around 0.35--0.50 mag/(100 d), except for the $r$ and $o$ bands.
In the late phases ($>300$d), a more rapid decline is observed in the lightcurves (1.15--1.30 mag/(100 d)), i.e., a break in the lightcurve. The transition is not well captured in \autoref{fig:lightcurve} due to data gaps in the observations. With straight line extrapolation in the data gap, we determine the lightcurve break to be at roughly $\sim$ day 300. The implication of this is discussed below. 

\begin{figure}
    \centering
    \includegraphics[width=\linewidth]{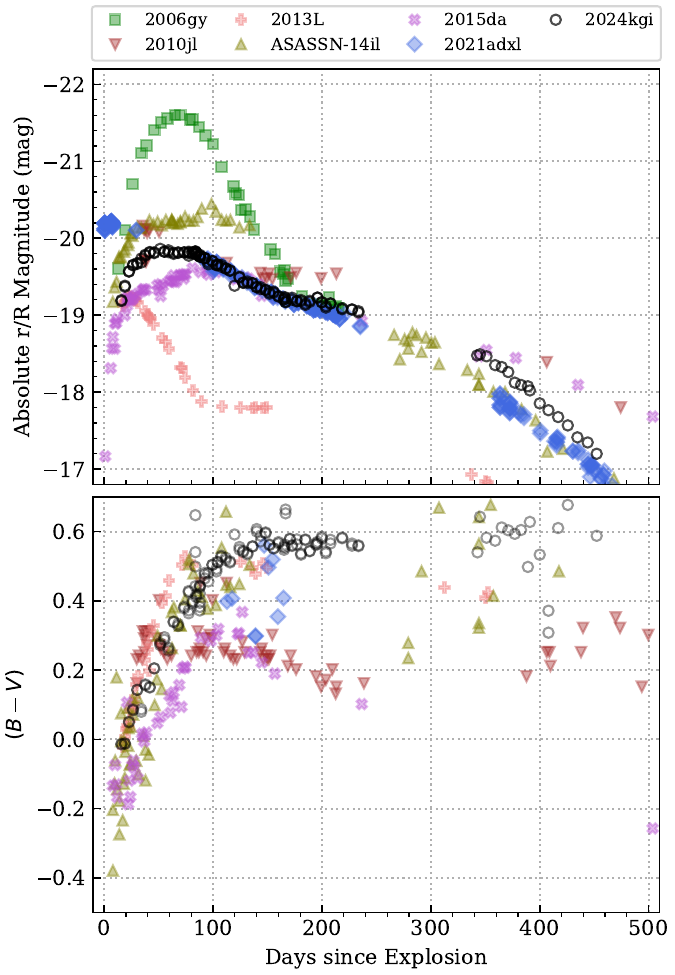}
    \caption{\textit{Top} panel shows the absolute $r/R$ band lightcurves of SN~2024kgi and the comparison SNe~IIn. \textit{Bottom} panel shows the intrinsic {\it B-V} color evolution of the SNe.}
    \label{fig:lk_comparison}
\end{figure}

We compare the lightcurve of SN~2024kgi with those of other well-studied SNe to emphasize its properties. As comparison objects, we include SNe~2006gy \citep{2006gy_smith_mccray_shellshockeddiffusion, 2006gy_agnoletto}, 2010jl \citep{2010jl_stoll, 2010jl_dust_smith, 2010jl_fransson, 2010jl_jencson, 2010jl_tsevtkov}, 2013L \citep{2013L_andrews, 2013L_taddia}, ASASSN-14il \citep{asassn14il_dukiya}, 2015da \citep{2015da_tartaglia, 2015da_smith} and 2021adxl \citep{2021adxl_brennan, 2021adxl_salmaso}, all of them with very luminous and long-lasting lightcurves, and showing strong Balmer lines throughout their evolution. We note that the explosion epoch of SNe~2010jl and 2021adxl is not well constrained; we adopt the time of first detection \citep{2010jl_stoll, 2021adxl_brennan} as the explosion epoch throughout this paper.
\autoref{fig:lk_comparison} shows the absolute $r$/$R$ band lightcurves, and the intrinsic {\it B-V} color evolution of this SN sample. SN~2024kgi exhibits a luminosity comparable to those of SNe~2010jl, 2015da, 2021adxl, and ASASSN-14il. The flattening of the lightcurve in SN~2024kgi is evident (in contrast with SN~2021adxl). A similar $r$-band lightcurve flattening is seen in SN~2010jl, and was attributed to continued strong interaction with the CSM, finding supported by the persistent Balmer emission lines \citep{2010jl_zhang}.
The late-time lightcurve slope of SN~2024kgi is similar to those observed in SN~2021adxl and ASASSN-14il. 

Another similarity shared by SNe~2010jl and 2024kgi is a break in the optical lightcurves after which the decline drastically steepens. A similar break is also seen in ASASSN-14il, SNe~2015da, and 2021adxl, where the decline rates increase significantly after a plateau at earlier phases. In the case of SN~2010jl, \cite{2010jl_ofek2014} considered three possible scenarios to explain the break in the lightcurve - (i) The mass of the accumulated CSM at the shock front reaches the mass of the ejecta, i.e., a snow-plow phase; (ii) The shock becomes slow-cooling if the density is low enough, i.e. Sedov-Taylor phase; (iii) Alternatively, the shock reaches the end of dense CSM. Based on the data, they deem the first option most plausible; however, \cite{Moriya_2010jl_snow_plow} argues that the break is likely due to the shock reaching the end of the dense CSM. Based on their lightcurve modeling, \cite{Moriya_2010jl_snow_plow} find that the transition into the snow-plow phase is much more gradual than the observed timescale. However, these arguments predict a break in the total bolometric luminosity. Later NIR lightcurves of SN~2010jl presented by \cite{2010jl_fransson} show a drastic increase, following the optical break. Consequently, the break in the total pseudo-bolometric luminosity (over a wavelength range of 3600--24000~\AA) is less pronounced. Most likely, the reprocessing of the optical luminosity into NIR photons by newly formed dust in a CDS and, eventually, in the SN ejecta \citep{2010jl_dust_smith, 2010jl_maeda_2013, 2010jl_gall_2014, 2010jl_Sarangi_2018} plays a significant role in the observed optical break in SNe~2010jl (and, possibly, also in SN~2024kgi). We discuss this further in \autoref{sec:bolometric}.

The intrinsic {\it B-V} color evolution of SN~2024kgi along with the comparison sample is also shown in \autoref{fig:lk_comparison}. We see that the color of all the SNe shows a very quick redward evolution as their photosphere cools down. Later, we see that the color of SN~2024kgi becomes roughly flat, matching the evolution of SN~2013L and ASASSN-14il. Interestingly, the color evolution of SNe~2010jl and 2015da shows a digression from this behaviour as their color shows a gradual blueward evolution, but it should be noted that the strong \hb line in the V-band for these SNe may affect the color evolution.

\begin{table}
    \centering
    \begin{tabular}{cccccc}
\hline \hline 
 & Rise Time & \multicolumn{2}{c}{Peak Magnitude} \\
 &  & Observed & Absolute \\
 & (day) & (mag) & (mag) \\
\hline
U    & $41.3^{+2.1}_{-3.5}$ & $16.78^{+0.01}_{-0.01}$ &  $-19.56 \pm 0.06 $\\
B    & $41.2^{+2.1}_{-3.2}$ & $16.78^{+0.01}_{-0.01}$ &  $-19.54 \pm 0.06$\\
g    & $44.6^{+1.7}_{-2.7}$ & $16.55^{+0.01}_{-0.01}$ &  $-19.70 \pm 0.06$\\
V    & $52.5^{+1.4}_{-2.3}$ & $16.48^{+0.01}_{-0.01}$ &  $-19.81 \pm 0.06$\\
r    & $58.2^{+3.9}_{-2.7}$ & $16.38^{+0.01}_{-0.01}$ &  $-19.83 \pm 0.06$\\
o    & $54.1^{+10.1}_{-4.5}$ & $16.35^{+0.01}_{-0.02}$ &  $-19.84 \pm 0.06$\\
i    & $74.8^{+2.2}_{-3.8}$ & $16.36^{+0.01}_{-0.01}$ &  $-19.80 \pm 0.06$\\
\hline \hline 
 & \multicolumn{2}{c}{Decline Rate} \\
 &  Day 100--150 & 170--230 & 330--450\\
 & (mag/100d) & (mag/100d) & (mag/100d) \\
\hline
U    & $1.57 \pm 0.09$ & $0.49 \pm 0.11$ & $-$ \\
B    & $1.15 \pm 0.03$ & $0.41 \pm 0.03$ & $1.29 \pm 0.04$ \\
g    & $1.08 \pm 0.05$ & $0.35 \pm 0.05$ & $1.20 \pm 0.02$ \\
V    & $0.95 \pm 0.03$ & $0.42 \pm 0.02$ & $1.27 \pm 0.04$ \\
r    & $0.76 \pm 0.03$ & $0.20 \pm 0.05$ & $1.15 \pm 0.03$ \\
o    & $0.85 \pm 0.02$ & $0.34 \pm 0.03$ & $1.16 \pm 0.05$ \\
i    & $0.93 \pm 0.03$ & $0.52 \pm 0.03$ & $1.29 \pm 0.03$ \\
\hline
\end{tabular}

    \caption{Rise times, observed and absolute peak magnitudes, and decline rates at distinct phases of the lightcurve evolution of SN~2024kgi.}
    \label{tab:peak_rise_slope}
\end{table}

\section{Bolometric Lightcurve} \label{sec:bolometric}
\begin{figure}
    \centering
    \includegraphics[width=1.\linewidth]{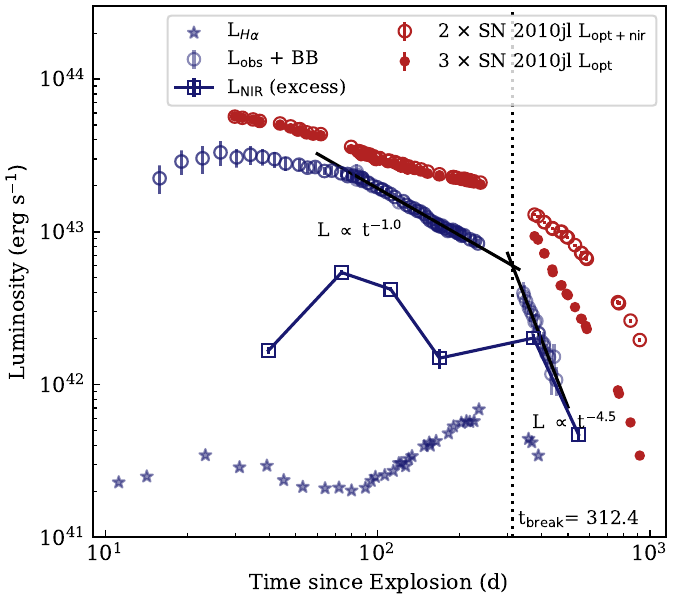}
    \caption{Evolution of the bolometric luminosity of SN~2024kgi. The bolometric lightcurve is best fitted by two power-laws, with an estimated break time at day $\sim 312$, indicated by the vertical dotted line. Lightcurve of excess NIR flux estimated from secondary blackbody fit to the SED, and evolution of \ha luminosity, are also shown. For comparison, we plot the observed optical and optical+NIR pseudo-bolometric luminosity of SN~2010jl, which is multiplied by an offset factor for clarity.}
    \label{fig:bolometric}
\end{figure}

We construct the bolometric lightcurve using {\texttt Superbol} \citep{2018Nicholl}. The unreddened $UBgcVi$ lightcurves are converted to fluxes and integrated to estimate the observed luminosity. We exclude the $r$ and $o$ bands, as they contain a significant contribution from the strong H$\alpha$ emission line. We estimate the contribution from radiation outside the observed wavelength range using a bolometric correction calculated by fitting a blackbody at each epoch. At early times ($< 300$ d), the omission of the $r$ and $o$ bands have a negligible effect on the observed bolometric luminosity and the correction factor. However, at later times, including $r$ and $o$ bands, raises the correction factor significantly, due to the flux excess.
We do not account for any NIR excess (see \autoref{sec:nir_spec}) in the bolometric correction.
The bolometric lightcurve is plotted in \autoref{fig:bolometric}, along with the quasi-bolometric (Optical and Optical+NIR) lightcurves of SN~2010jl for comparison. The bolometric lightcurve peaks at $\sim 3 \times 10^{43}$ erg s$^{-1}$ at around $\sim 30$ days from explosion, even earlier than the $U$ band peak. This indicates that a significant portion of the luminosity is emitted in UV at early times, which dominates the bolometric lightcurve. We calculate the total radiated energy at optical wavelengths during our monitoring campaign by integrating over the time the observed bolometric lightcurve: We obtain $\sim 2 \times 10^{50}$ ergs. This matches the median radiated energy for the slow-luminous group of \cite{hiramatsu_IIn_optical}. After maximum, the lightcurve settled onto a power-law decline (with L $\propto$ t$^{-1.0}$) up to day $\sim 240$, after which the object hides behind the Sun. When the object became visible again, the lightcurve behavior dramatically changed, exhibiting a steeper decline (L $\propto$ t$^{-4.5}$). We estimate the transition to occur on day $\sim 312$ by extrapolating the power-law fits from both regions.

The optical quasi-bolometric lightcurve of SN~2010jl also shows a significant break at a similar timeframe; however, the severity of the break is considerably reduced by including the contribution of the NIR bands in the bolometric lightcurve, as they contain a significant portion of the total flux. A significant amount of new dust is predicted to form at this epoch from the NIR excess and extincted H-line profiles \citep{2010jl_maeda_2013, 2010jl_Sarangi_2018}. Dust formation reprocesses optical photons into NIR flux and causes the faster decline of the optical lightcurves. New dust formation is also expected in SN~2024kgi, after the lightcurve break (see \autoref{sec:discussion_dust_formation}), as this could account for a fraction of the flux deficit in the optical domain. We lack NIR photometry for SN~2024kgi to verify this statement. However, we have fitted double blackbodies to the SED of SN~2024kgi (see \autoref{sec:nir_spec}). We integrate the secondary blackbody to calculate a sparse NIR lightcurve, which is presented in \autoref{fig:bolometric}. We note a rise in the NIR lightcurves after the break, similar to SN~2010jl. Additionally, we see that the NIR excess is almost an order-of-magnitude smaller than the presented bolometric lightcurve before the optical break. However, after the optical break, the NIR excess becomes significant, and therefore the bolometric luminosity is underestimated.

Nonetheless, even the total (Optical+NIR) bolometric lightcurve of SN~2010jl shows a break. When the ejecta interacts with CSM with a power-law density structure ($\rho_{csm} \propto r^{-s}$), the power deposited by the forward shock follows a power-law behavior as well \citep{CW12, takashi_IIn_analytical, 2010jl_ofek2014}. Therefore, a change in the power-law index of the lightcurve signifies either a change in composition or termination of the forward shock as the shock emerges from the CSM. Alternatively, explanations such as a transition into a snow-plow phase have been proposed \citep{2010jl_ofek2014}, but the accumulation of enough swept-up CSM mass requires significantly longer timescales \citep{Moriya_2010jl_snow_plow}.

We also calculate the \ha luminosity contribution after the subtraction of the spectral continuum near the \ha region. The evolution of \ha luminosity is plotted in \autoref{fig:bolometric}. The early H$\alpha$ luminosity primarily sees contribution from the narrow component and the Lorentzian wings. The narrow component, powered by excitation from the continuum emission, rises and decays in tandem with the bolometric lightcurve \citep{2006tf_smith}. Later, as the intermediate and broad components emerge, the \ha lightcurve increases consistently after day $\sim 90$, contrary to the bolometric lightcurve. The \ha luminosity is primarily powered by the ongoing interaction with the CSM. Later, the \ha lightcurve experiences a drastic break, at a timescale similar to the optical break, again indicating a change in the CSM structure. A similar evolution of the \ha luminosity is also seen in SN~2015da \citep{2015da_tartaglia, 2015da_chugai}.

There have been many attempts to explain the behavior of bolometric lightcurves of these interacting transients. The ejecta and CSM usually follow a power-law density structure ($\rho_{ej} \propto r^{-n}$, $\rho_{csm} \propto r^{-s}$).
\cite{Khatami_Kasen_2024} provides a framework to classify interacting transients in four distinct categories that have unique lightcurve characteristics based on the combination of two parameters -- the CSM to ejecta mass ratio ($\eta$), and the ratio of the diffusion timescale to the dynamical timescale ($\xi$). These parameters govern the shock deceleration and location of the shock breakout, respectively.
The lightcurve properties of interacting transients are then classified in the following categories -- (i) Edge breakout - light CSM ($\xi \gg 1, \eta \ll 1$), (ii) Edge breakout - heavy CSM ($\xi \gg 1, \eta \gtrsim 1$), (iii) Interior breakout - light CSM ($\xi \lesssim 1, \eta \ll 1$), (iv) Interior breakout - heavy CSM ($\xi \lesssim 1, \eta \gtrsim 1$).
By comparing the lightcurve shape, we can infer that SN~2024kgi experiences an interior shock breakout ($\xi \lesssim 1$). 
Transients experiencing interior shock breakout in the CSM show a prominent continued interaction phase, where the photons generated by the forward shock almost instantaneously escape and therefore the observed luminosity traces the forward shock luminosity \citep{takashi_IIn_analytical, Tsuna_IIn_model_2019}. The analytical relation between the lightcurve power-law index ($\alpha$; $L \propto t^{\alpha}$) and $n$ \& $s$ by \cite{Khatami_Kasen_2024} for the continued interaction comes out to be $\alpha = (2n + 6s - ns - 15)/(n - s)$, which is the same as other works in this regime \citep{CW12, takashi_IIn_analytical, 2010jl_ofek2014}.

In the case of SN~2024kgi, we observe the luminosity follows a $t^{-1.0}$, which corresponds to a CSM density index of $3$, much steeper than expected from a steady mass loss. This indicates that the mass-loss rates increased prior to the SN explosion. We note this derived time dependence (and by extension, the CSM density index) is sensitive to the explosion epoch. Our derived explosion time corresponds to the epoch when photons start to escape from the CSM; the CSM-ejecta interaction may start much sooner in the case of an optically thick CSM.

\subsection{Modeling the bolometric lightcurve}
As previously stated, there are various approaches to model the lightcurve under different assumptions. \cite{Chevalier_1982_self-similar} presented self-similar solution to explain the hydrodynamics of ejecta ($\rho_{ej} \propto r^{-n}$) interacting with CSM material ($\rho_{csm} \propto r^{-s}$). \cite{CW12} builds upon this solution and gives a general analytical expression for the expected instantaneous luminosity and provides a generalized diffusion scheme expanding the work of \cite{Arnett_1980, Arnett_1982}. However, this model assumes that the diffusion time is constant throughout the evolution, which can be severely incorrect in some cases, especially at late phases when the shock is at low optical depths, and photons escape quickly. Several authors have studied the opposite regime, in which photon escape times are almost instantaneous at late phases \citep{takashi_IIn_analytical, 2010jl_ofek2014, Tsuna_IIn_model_2019}. In this regime, the lightcurve traces the instantaneous energy deposited by the forward shock and therefore assumes a power-law shape.

The difficulty in accurately reproducing the lightcurve throughout the entire timescale lies in the energy deposited by a moving shock front and the constantly decreasing optical depths. This is properly taken into account in radiative hydrodynamic simulations. \cite{Transfit-CSM_zhang_2026} addressed this by numerically solving the mass-momentum equation and the diffusion equation simultaneously, to provide self-consistent lightcurves of interacting transients.

Here, we propose a model to analytically treat the variable diffusion timescale. We adopt the self-similar hydrodynamic solutions presented in \cite{Chevalier_1982_self-similar} and \cite{CW12}, which are summarized in \autoref{sec:chevalier_solutions}. These solutions provide us with the instantaneous shock luminosity deposited by the forward and reverse shocks, $L_{fs}$ and $L_{rs}$, respectively. The forward (and reverse) luminosities evolve as a power-law ($L_{fs/rs} \propto t^{\alpha}$), until the forward shock runs out of the dense CSM (or the reverse shock runs out of the ejecta).

Now, neglecting the adiabatic expansion, we have the diffusion equation as 
\begin{equation}
    \frac{\partial E(r,t)}{\partial m} = \varepsilon_{sh}(r, t) - \frac{\partial L}{\partial m}, \label{eq:diffusion_pde}
\end{equation}

\noindent
where $L(r,t)$ is the luminosity and $\varepsilon_{sh}(r,t)$ is the shock-heating rate. We can reduce it to a one-zone model by integrating over the mass coordinates.

\begin{equation}
    \frac{dE}{dt} = L_{inp} - L_{obs},
\end{equation}

\noindent
where E(t) is the energy of the central reservoir, $L_{inp}$ is the input luminosity from the forward and reverse shocks, and $L_{obs}$ is the observed luminosity. We can approximate $L_{obs}$ by assuming the energy diffuses over a characteristic timescale $t_d(t)$. Therefore,

\begin{equation}
    \frac{dE}{dt} = L_{inp} - \frac{E}{t_d(t)}  \label{eq:diffusion_ivp}.
\end{equation}

We can obtain an analytical expression for $E(t)$ by introducing an integrating factor $e^{\phi (t)}$, where

\begin{equation}
    \phi(t) = \int_0^t \frac{dt'}{t_d(t')}
\end{equation}

From this, we infer

\begin{equation}
    L_{obs}(t) = \frac{1}{t_d(t)} E(t) = \frac{e^{-\phi(t)}}{t_d(t)} \int_0^t L_{inp}(t') e^{\phi(t')} dt' \label{eq:Lobs_var_tdiff}
\end{equation}

In case of constant diffusion time, \autoref{eq:Lobs_var_tdiff} reduces to the solution given by \cite{CW12}.
\begin{equation}
L_{obs}(t) = \frac{1}{t_d} e^{-\frac{t}{td}} \int_0^{t} e^{\frac{t'}{td}} L_{inp}(t') dt' \label{eq:Lobs_fix_tdiff}
\end{equation}

Conceptually, \autoref{eq:Lobs_var_tdiff} can be thought of as a collection of delta pulses diffusing through media with different diffusion times. A similar one-zone approach for the moving shock front is adopted by \cite{Sarin_redback_interacrting}. Further, we can account for the moving shock front by introducing a delay factor in the input luminosity corresponding to the light travel time from the shock front ($R_sh$) to the outer edge of the CSM ($R_{csm,out}$).

\begin{equation}
\begin{split}
    L_{obs}(t) = \frac{e^{-\phi(t)}}{t_d(t)} \int_0^t L_{inp}(t' - \frac{\Delta R}{c}) e^{\phi(t')} dt' \label{eq:Lobs_time_delay}
\end{split}
\end{equation}

\noindent
where $\Delta R = (R_{csm,out} - R_{sh})$. \autoref{eq:Lobs_time_delay} gives the observed luminosity as a function of time. However, it can become numerically unstable for diminishing $t_d$, so equivalently \autoref{eq:diffusion_ivp} can be solved as an initial value problem.

We evaluate the time-variable diffusion time $t_d(t)$ as 

\begin{equation}
    t_d(t) = \frac{1}{c} \int_{r1}^{r2} \tau(r)dr \label{eq:diffusion_time},
\end{equation}

\noindent
where $r1$ and $r2$ define the boundary of the diffusion region. In case of forward shock, $r1 = R_{fs}$ and $r2 = R_{csm,out}$. $\tau(r)$ is the optical depth of diffusion mass.

\begin{equation}
    \tau (r) = \int_{r1}^{r2} \kappa \rho(r) dr,
\end{equation}
where $\kappa$ is the gray opacity of the diffusing material.

Here, we discuss the caveats of our model. We assume a radiation-dominated system and neglect adiabatic losses, which become relevant near and after the shock emergence. We assume a time-constant gray opacity and a constant conversion efficiency of kinetic energy into radiation. Both quantities can vary with the density, temperature, and composition of the material. Specifically, the conversion efficiency can be higher in a denser CSM at the beginning of the interaction but can drop significantly near shock emergence, as a substantial amount of energy may be released in UV and X-ray radiation. Additionally, since \cite{Chevalier_1982_self-similar} solution are derived for 1D system, effects of asymmetry in the CSM can not be modeled.

\begin{figure}
    \centering
    \includegraphics[width=\linewidth]{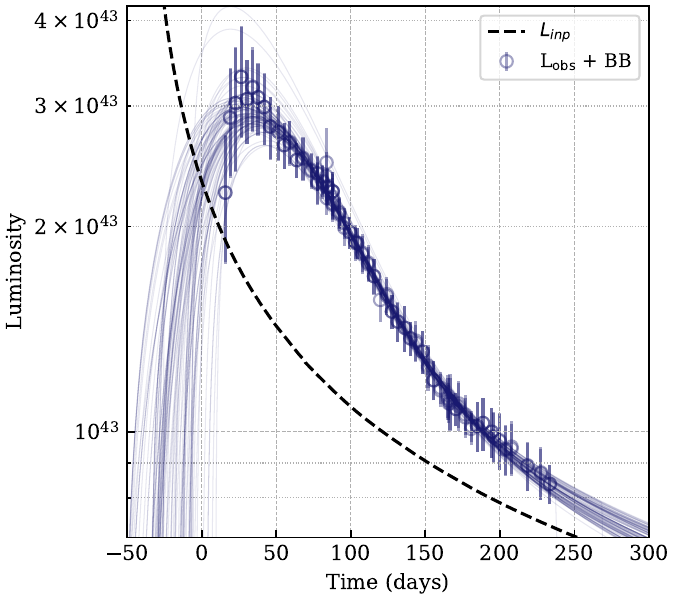}
    \includegraphics[width=\linewidth]{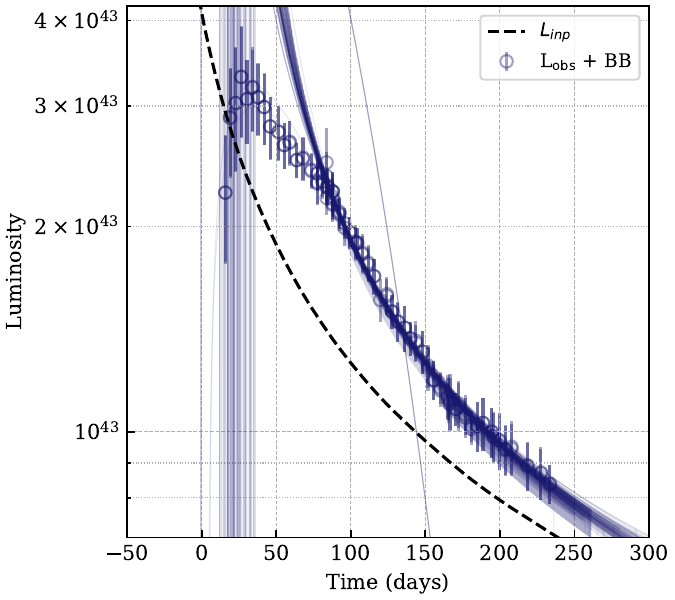}
    \caption{The modeled lightcurves are plotted along with the observed bolometric luminosity for Model 1 (\textit{top}) and Model 2 (\textit{bottom}). Each line corresponds to a randomly sampled parameter vector from the posterior, and the bold line represents the median of all such lines. The dashed black line shows the corresponding median instantaneous power input from the forward shock.}
    \label{fig:bolometric_model}
\end{figure}

We adopt this scheme to obtain the observed forward and reverse shock luminosities, and combine them to compute the total observed luminosity for a given set of parameters. We use this model to fit the bolometric lightcurve of SN~2024kgi with MCMC using \texttt{emcee} \citep{emcee_Foreman-Mackey_2013} to sample the posterior. We only fit the lightcurve before the break, as once the forward shock terminates, the assumption of a stationary photosphere is no longer valid. We fix the power-law index of the outer ejecta $n=10$ \citep{matzner_mckee_ibc_ejecta_density} and the opacity $k=0.34$~cm$^2$ g$^{-1}$ for fully ionized solar composition material. We take the remaining parameters as fit parameters, the priors of which are mentioned in \autoref{tab:bolometric_params}, along with the estimated parameter values. We derive two sets of inferred parameters by -- (i) including the peak of the lightcurve in the fit (Model 1), (ii) fitting the lightcurve after day 80, when the lightcurve has settled into a power-law decline (Model 2). The fitted models are plotted in \autoref{fig:bolometric_model}, and both models reproduce the post-peak luminosity accurately. 

\begin{table*}
    \caption{The priors and fitted parameters for the bolometric lightcurve model.}
    \label{tab:bolometric_params}
    \centering
    \begin{tabular}{ccccccc}
    \hline\hline
        Parameter & M$_{\mathrm{ej}}$ & v$_{\mathrm{ej}}$ & s & M$_{\mathrm{csm}}$ & R$_{csm,in}$ & log $\rho_{in}$ \\ 
        Unit & (M$_{\odot}$) & (km s$^{-1}$) & -- & (M$_{\odot}$) & (AU) & (g cm$^{-3}$) \\ 
        Prior & -- & $2\times 10^3$ -- $2\times 10^4$ & 0--2.9 & 0--100 & 1--500 & -14 -- -9 \\ 
        \hline
        \hline \\
        Model 1 & $8.70^{+32.97}_{-6.76}$ & $3006^{+871}_{-870}$ & $2.39^{+0.04}_{-0.05}$ & $34.7^{+13.2}_{-7.1}$ & $10.5^{+51.2}_{-8.2}$ & $-10.4^{+1.5}_{-1.6}$ \\ 
        \\
        Model 2 & $8.51^{+32.22}_{-6.46}$ & $9774^{+3853}_{-3516}$ & $2.74^{+0.07}_{-0.10}$ & $2.34^{+11.8}_{-1.6}$ & $9.3^{+36.4}_{-9.3}$ & $-11.0^{+1.7}_{-1.8}$ \\
        \\
    \hline
    \end{tabular}
\end{table*}


Model 1 reproduces the peak lightcurve well, but the rise of the lightcurve seems to be steeper than the model, and the peak luminosity is underpredicted. However, these discrepencies are roughly within the large error-bars due to uncertainty in the early time bolometric correction factor. The CSM mass predicted by Model 1 is very high ($34.7^{+13.2}_{-7.1} \, M_{\odot}$), and therefore requires mass-loss rates of the order $\sim 1 \, M_{\odot}$ (see \autoref{sec:discussion_mass_loss}. Additionally, the ejecta velocities predicted by Model 1 ($\sim 3000$ km s$^{-1}$) are at odds with the observed ejecta velocities ($\sim 10,000$ km s$^{-1}$; see \autoref{sec:spectral_evolution}).
Model 2 predicts peak luminosities much higher than the observed one, but can explain the post-peak luminosity with a modest amount of CSM ($2.34^{+11.8}_{-1.6} \, M_{\odot}$). The inferred ejecta velocities of ($\sim 10000$ km s$^{-1}$) are in agreement with the spectroscopic measurements. 

We note that the ejecta velocities and CSM mass can show degeneracy \citep{takashi_IIn_analytical}, such that higher ejecta velocities can produce similar luminosity with lower CSM mass. This is the fundamental difference between Models 1 and 2. Model 1 fits the initial part of lightcurve much better, but requires massive mass-loss rates and ejecta velocities that disagree with the spectroscopic measurements. The high CSM mass in Model 1 corresponds to high optical depths, smoothing the breakout and accurately reproducing the peak luminosity. Model 2 can faithfully reproduce the post 80 day lightcurve with a comparatively modest amount of CSM combined with higher ejecta velocities. However, because peak luminosity and rise time are not fitted in Model 2, the ejecta velocities and CSM mass have larger uncertainties. We also note that our model assumes spherical symmetry, and lightcurve properties like rise time and peak luminosity will be influenced by variations in the diffusion timescale caused by any asymmetry, as is the case.
Taking into consideration the inferred ejecta velocities and the much more feasible mass-loss rates, Model 2 seems more plausible than Model 1, even though the latter fits the lightcurve better, though we discuss the implications of both in \autoref{sec:discussion_mass_loss}.
The poor match between Model 2 and the bolometric lightcurve could be due to a plethora of reasons, including deviations from a strictly power-law CSM structure, energy released outside the optical wavelength that could not be properly accounted for by bolometric corrections, and all the caveats discussed earlier. 

The value of $s$ directly influences the shape of post-peak lightcurve, we have good constraints on the CSM density exponent from both models. Both models predict values of $s > 2$, indicating deviations from a steady mass-loss. The predicted values of $s$ (Model 1 -- $2.39^{+0.04}_{-0.05}$, Model 2 -- $2.74^{+0.07}_{-0.10}$) deviate from the $s \sim 3$ derived by using the analytical scaling of \cite{takashi_IIn_analytical}, and \cite{Khatami_Kasen_2024}. However, after inspecting \autoref{fig:bolometric_model}, the reason is immediately clear. \cite{takashi_IIn_analytical}, and \cite{Khatami_Kasen_2024} assume that the luminosity is directly powered by the forward shock. However, in this case, the CSM shell is so massive and extensive that even at day 100--200, the optical depth are still very high, and the lightcurve still has a significant contribution from energy injected at earlier times. \cite{Transfit-CSM_zhang_2026} also finds a similar lightcurve morphology for extended CSM structures. This effect can seemingly produce a steeper power-law decline than expected from a directly shock-powered case.

\section{Spectroscopic Evolution} \label{sec:spectral_evolution}

\begin{figure*}
    \centering
    \includegraphics[width=0.97\textwidth]{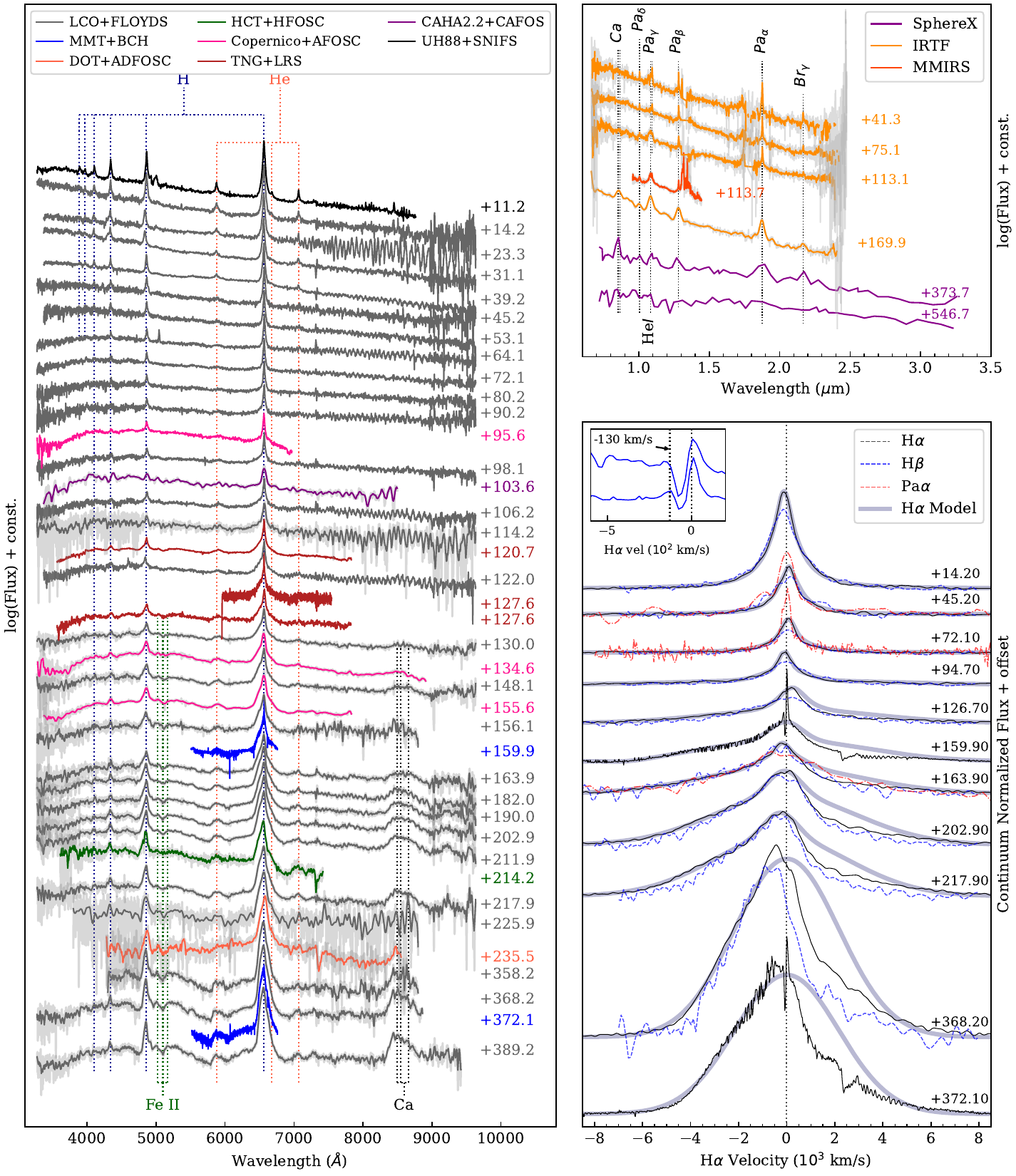}
    \caption{\textit{Left} panel shows optical spectral evolution of SN~2024kgi. Spectra with a larger noise (shown in light gray), are smoothed with a Savitsky-Golay filter. \textit{Top right} panel shows the NIR spectral evolution of SN~2024kgi. The gray lines show the original spectrum, and the colored lines show the smoothed spectrum. \textit{Bottom right} panel shows the evolution of continuum-subtracted \ha lines. \hb and \pa lines are overplotted to show that they match the unattenuated left wing of H$\alpha$. The gray lines represent the model fit to the H$\alpha$ profile. The inset shows the narrow component of \ha, which is resolved in the higher resolution MMT spectra.}
    \label{fig:spectra}
\end{figure*}

\autoref{fig:spectra} shows the spectral evolution of SN~2024kgi from day +11 to day +389. The early spectra show a blue ($\sim 10,000$~K) continuum and are dominated by narrow H and He lines with Lorentzian wings, characteristic of high electron scattering optical depths. The \ha luminosity peaks on a similar timescale to the bolometric lightcurve and then starts to decline (see \autoref{fig:bolometric}).
Later, by day $\sim +90$, the spectral continuum cools to $\sim 6000$~K. 
From day +98, the ejecta signature begins to emerge as broad H line components. Broad \ion{He}{1} $5876$ \AA, He $6678$ \AA and the \ion{Ca}{2} triplet ($8498, 8542, 8662$ \AA) are visible soon after day around +120. The H$\alpha$ line shows a composite multicomponent structure at these epochs due to contributions from the CSM ahead of shock, post-shock region, and freely expanding ejecta. The \ha luminosity starts to increase again after day $\sim 90$ due to the contribution from the CDS and the freely expanding ejecta, after declining initially.
In a spherically symmetric CSM configuration, we do not expect ejecta signatures to be revealed alongside the CDS features so early after maximum \citep{Dessart_IIn_simulation, smith_IIn_review}. Therefore, if the CSM deviates from spherical symmetry, the ejecta and the CDS may become visible simultaneously as the photosphere recedes from the unshocked CSM. This can be achieved in an axisymmetric CSM structure, such as a torus or a bipolar nebula.

The intermediate and broad features become more prominent over time as the continuum luminosity recedes. After day $\sim +170$, the spectral continuum gradually dilutes into blends of metal lines. The Balmer lines and \ion{Ca}{2} triplet continue to strengthen over this period. 

\begin{figure}
    \centering
    \includegraphics[width=\linewidth]{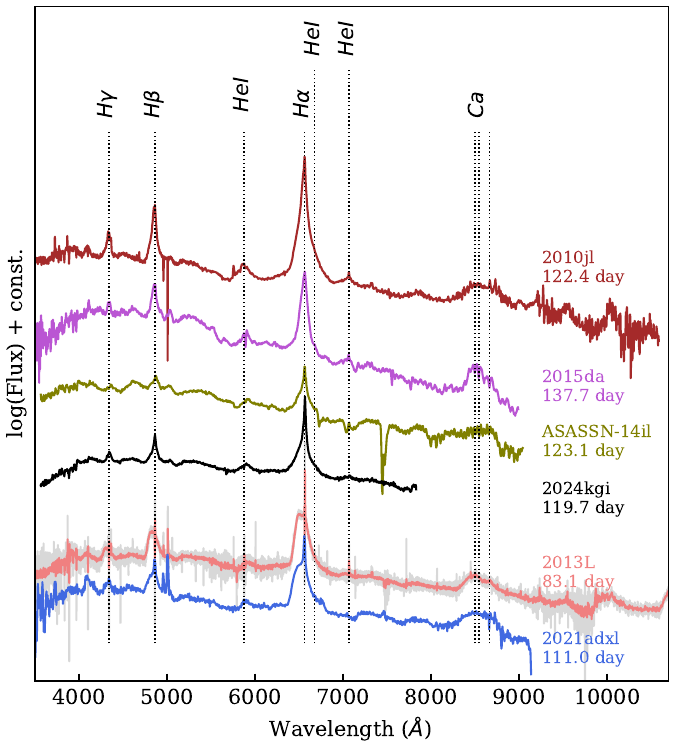}
    \includegraphics[width=\linewidth]{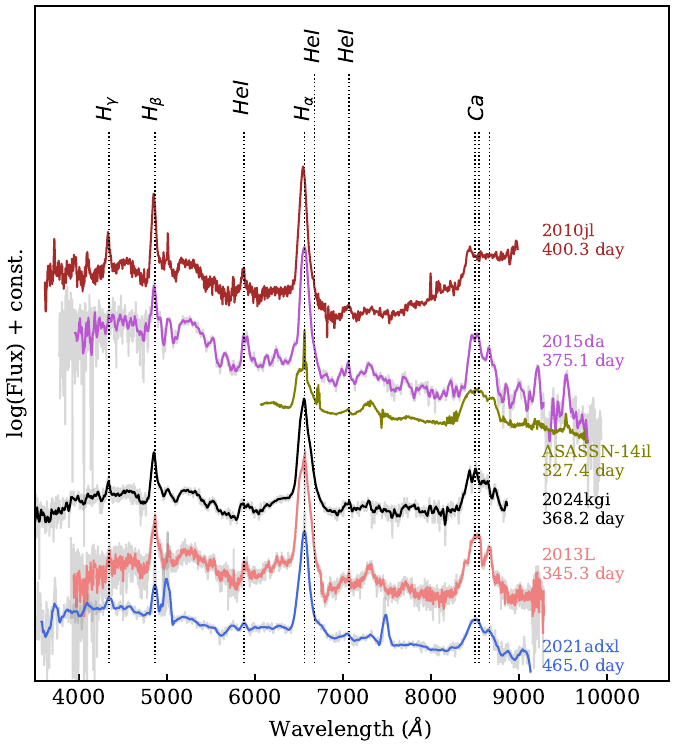}
    \caption{Spectral comparison of SN~2024kgi with the sample of SNe~IIn in the post-peak phase (\textit{top panel}), and at late phases (day $>+300$; \textit{bottom panel}).}
    \label{fig:spectra_comp}
\end{figure}

\begin{figure}
    \centering
    \includegraphics[width=\linewidth]{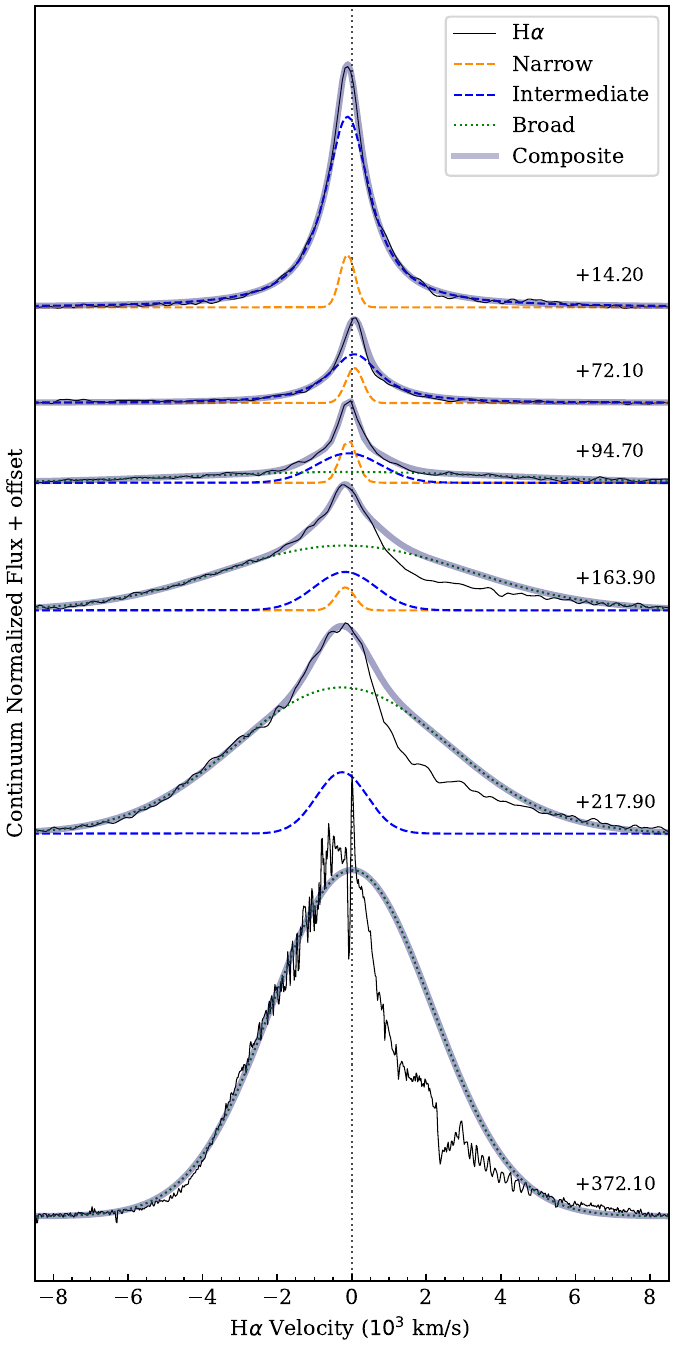}
    \caption{Multicomponent fits to the continuum normalized \ha profile at different stages of evolution. Only velocities bluer than 500 km s\power{-1} are fitted.}
    \label{fig:halpha_fit}
\end{figure}

In \autoref{fig:spectra_comp}, we compare the spectra of SN~2024kgi with those of SNe~IIn at two different epochs: after the peak ($\sim +100$ d) and at late phases ($> +300$ d). At $\sim +100$ days, all SNe have a strong blackbody-like continuum with prominent narrow H lines. 
The most striking feature noticeable is the presence of two types of H$\alpha$ profiles - i) a flat-topped H$\alpha$ profile with the evident red-side flux deficit seen in the spectra of SNe~2013L and 2021adxl; ii) a multicomponent H$\alpha$ profile with a comparatively subtle (but still noticeable) red-side flux deficit seen in SNe~2010jl, ASASSN-14il, 2015da, and 2024kgi. These two H$\alpha$ profile shapes can be correlated with distinct physical geometries of CSM. \cite{2013L_taddia} and \cite{2021adxl_brennan} argue the H$\alpha$ profiles exhibited by SNe~2013L and 2021adxl can be explained within the confines of spherical symmetry.
The boxy profile is a characteristic of a geometrically thin emitting region. And the flux deficit in the red side of the line can be attributed to the occultation by the continuum photosphere.  
However, an asymmetric disc/torus-like CSM structure, when viewed edge-on, can produce the same effect \citep{2013L_andrews}.

In contrast, the multicomponent H$\alpha$ profiles of SNe 2010jl, ASASSN-14il, and 2015da \citep{2010jl_jencson, 2010jl_zhang, asassn14il_dukiya, 2015da_tartaglia, 2015da_smith} are more comfortably explained by asymmetric CSM geometries, such as a disk/torus like or a bipolar CSM, which allow the ejecta to be detected at such early epochs. The red-sided flux deficit in the H$\alpha$ profiles can be explained as either occultation by the photosphere, extinction by newly formed dust in the post-shock gas, or a combination of both. 
SN~2024kgi shows striking similarity with the spectra of SNe~2010jl, ASASSN-14il, and 2015da. From the early emergence of ejecta signatures, we already expect SN~2024kgi to have an asymmetric CSM structure. However, if we assume the CSM to be axis-symmetric (such as a disk/torus or as a bipolar nebula CSM structure), it can naturally explain the observed deficit in the red-wing of H lines after the lightcurve peak.

At even ther late epoch ($>+300$ day), SNe~2010jl, 2013L, ASASSN-14il, 2015da, 2021adxl, and 2024kgi share many spectral features in common. The continuum photosphere dilutes into complicated blends of metal lines. The H and He lines have become more prominent and are dominated by intermediate and broad components. Strong \ion{Ca}{2} triplet features are observed in all SNe. A flux deficit in the red-wing of the H$\alpha$ profile is still present in the spectra of these SNe, but the emergence of the He I $6678$\AA\, line complicates a quantitative analysis. At this epoch, the blue-shifted H$\alpha$ profile of SNe~2010jl, 2015da, and ASASSN-14il is at least in part caused by new-dust formation in the post-shock gas and/or in the unshocked ejecta \citep{2010jl_maeda_2013, 2010jl_Chugai_2018, 2010jl_Sarangi_2018, asassn14il_dukiya, 2015da_smith}. Looking at the similarity of the H$\alpha$ profile with those observed in other SNe~IIn, dust formation may be a viable mechanism for the observed red-side flux deficit in SN~2024kgi. We further discuss this in detail in \autoref{sec:discussion_dust_formation}.

\subsection{H lines} 
\label{sec:h_lines}
We plot the evolution of the continuum-normalized H$\alpha$ lines in high S/N spectra of SN~2024kgi in the bottom right panel of \autoref{fig:spectra}. We also plot the continuum-normalized H$\beta$ and Pa$\alpha$ lines (at nearby epochs), scaled to visually match the blue-side flux of H$\alpha$. Since the source lies at the outskirts of the host galaxy, we do not expect any significant contribution from the host galaxy \ion{H}{2} regions. This assumption is also supported by the lack of narrow \ion{N}{2} and \ion{S}{2} lines in the spectra typically associated with contaminating sources. As mentioned in \autoref{sec:spectral_evolution}, the early H$\alpha$ profiles show prominent Lorentzian wings characteristic of high electron scattering optical depths. We can faithfully model the H$\alpha$ profile with a combination of a narrow Gaussian core and broad Lorentzian wings, up to day $\sim72$. The Lorentzian wings can be characterized by an FWHM of $1500-2000$ km s\power{-1}. The \hb and Pa$\alpha$ lines show similar characteristics. All H lines are also symmetric with respect to the rest velocity. At this epoch, the continuum photosphere is likely in the unshocked CSM, and the \ha photons from the CDS and ejecta are completely opaque to the observer. 

At intermediate epochs (day 80--250), we gradually start to see the high-velocity wings coming from the unshocked ejecta. At these epochs, the profiles of \ha lines are better represented as a combination of multiple Gaussian components. A characteristic feature of the H line profiles in this epoch is the flux deficit in the red-shifted wing, which gradually becomes prominent as the continuum fades from the unshocked ejecta and reveals the underlying material. This causes an asymmetry between the blue and the red wings of the line profiles between $\sim$ 500--5000 km s$^{-1}$. We fit the H$\alpha$ profile with a combination of narrow, intermediate, and broad components, indicating contributions from the unshocked CSM, the CDS, and the freely expanding ejecta, respectively. In our fit, we only include velocities bluer than -500 km s\power{-1}. For the broad Gaussian component, the FWHM declines from $\sim 10000$ km s\power{-1}\, to $\sim 6000$ km~s\power{-1}\, during this phase. The intermediate Gaussian component exhibits a FWHM of 1500--2000 km s\power{-1}. The blue side of \ha line is accurately reproduced by this model, and the difference between the model and red-shifted wing can be used to quantify the flux deficit. The continuum subtracted \ha profiles at a few key epochs, along with the different components of the fit, are shown in \autoref{fig:halpha_fit}.

After comparing the \ha line to \hb and \pa, we can see that the flux deficit is not wavelength-dependent at these epochs and, therefore, not the consequence of new dust formation in the post-shock gas or into the ejecta. Another cause for the asymmetry could be a CSM structure with asymmetries in the redder and bluer sides (hence, a deviation from an axis-symmetric CSM), such that the CSM at the observer's side is either denser or located closer to the explosion \citep[e.g., in  SN~2012ab;][]{2012ab_gangopadhyay}. However, such a CSM structure would cause asymmetries in the broad component as well, and we would expect higher ejecta velocities from the red-wings of the line profiles. But the \ha wings at $> 5000$ km s\power{-1} are roughly symmetric. The obscuration of the receding material by the continuum photosphere is a more reasonable explanation, as inferred for many luminous SNe~IIn \citep{2013L_taddia, 2010jl_dust_smith, 2015da_smith}. 

At later epochs, the \ha lines are much more prominent compared to the continuum luminosity. The flux deficit remains; however, we can now clearly see a wavelength dependence when comparing \ha with H$\beta$. The peak of the line profile is also blueshifted, indicating that the obscuration extends beyond the red wing, attenuating emission near zero velocity and part of the blue wing. Similar line profiles and a wavelength-dependent flux deficit are also seen in SNe~2010jl, 2015da, and ASASSN-14il at comparable epochs, likely due to newly formed dust in the post-shock gas and/or ejecta. The blue wing can now be approximately represented by a single broad component. This change is temporally coincident with the break in the lightcurve, and supports the assumption that the shock has run out of the dense CSM.

Although in most of our low-resolution spectra the narrow component of \ha is unresolved, the two higher-resolution spectra obtained with MMT+BCH resolve it. The narrow component for these two epochs is plotted in the inset of the bottom right panel of \autoref{fig:spectra}. We see a distinct P-Cygni profile characteristic of the CSM outflow. We estimate the CSM velocity from the blue edge of the P-Cygni profile to be $\sim 130$ km s\power{-1}.

\subsection{NIR spectra} 
\label{sec:nir_spec}
\autoref{fig:spectra} shows the NIR spectral evolution of SN~2024kgi from day $\sim 40$ to day $\sim ~546$. Similar to the optical spectral evolution, the NIR spectra are dominated by H (Paschen and Bracket spectral lines) and He features. Initially, the lines show narrow cores with extended Lorentzian wings. In tandem with the optical spectra, the lines evolve into multicomponent features. The spectra taken at days $\sim 170$ and $\sim 374$ have a lower resolution, but the prominent H lines can still be recognized, although unresolved. The last low-resolution spectrum taken at day $\sim 546$ has very poor S/N, and none of the expected features are discernible. 

We note that the optical+NIR continuum cannot be explained by a single blackbody component. Therefore, we study the evolution of the spectral continuum by fitting it with a 2-component model --- (i) an optical blackbody, (ii) Excess emission characterized by either a blackbody or optical thin dust. These components will hereafter be referred to as the hot and warm components. The emissions from the blackbody and optically thin dust have the following forms, respectively:

\begin{equation}
    F_{\lambda}(T, R) = \frac{\pi R^2}{D^2} B_{\lambda}(T),
\end{equation}
\begin{equation}
    F_{\lambda}(T_d, M_d) =  \frac{M_d \kappa(\lambda)}{D^2} B_{\lambda}(T_d),
\end{equation}

\noindent
where $B_{\lambda}$ is the Planck function, $R$ is the radius of the blackbody, $D$ is the luminosity distance to the source, and $T$ (or $T_d$) is the characteristic temperature of the blackbody (or dust), and $M_d$ is the total mass of the dust.

Due to the lack of NIR photometry, we match the overlapping region of the NIR spectra with nearby optical spectra to get an absolute flux calibration. For the last spectrum (day $\sim 547)$, we do not have a corresponding optical spectrum available, so we convert the optical photometry of nearby epochs into fluxes and use that for fitting. We note that this means the absolute flux calibration of the last NIR spectrum is uncertain. For the last epoch, we also fix the temperature of the hot blackbody component to the value obtained at the previous epoch (T $\sim 6300$ K). The fit is performed with MCMC methods using \texttt{emcee} \citep{emcee_Foreman-Mackey_2013}. The emission lines are masked and not included in the fit. Following \cite{Fox_spitzer_IIn_2011}, we adopt a graphite/silicate dust composition with a grain size of 0.01 $\mu m$ or 0.1 $\mu m$; the dust opacities are taken from \cite{Fox_2005ip_2010}. 

\begin{figure*}
    \centering
    \includegraphics[width=\linewidth]{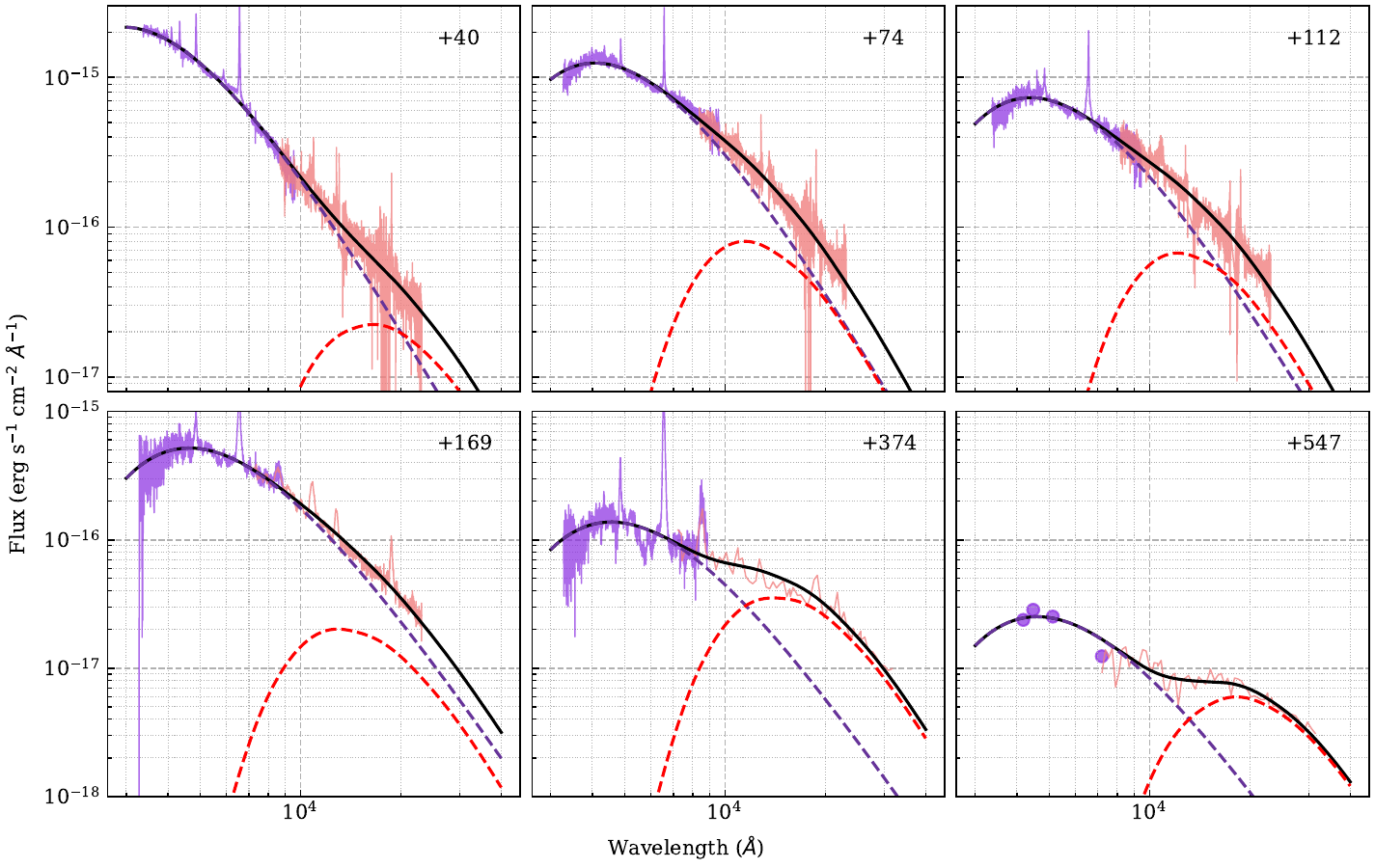}
    \caption{Combined optical and NIR spectra at nearby epochs. The multicomponent fit (Blackbody + 0.1 $\mu$m graphite dust) is shown as a solid black line, and the contribution from the individual components is shown with dashed lines.}
    \label{fig:bb_fit}
\end{figure*}

The combined spectra along with the fitted model for dust emission from 0.1 $\mu m$ graphite dust are shown in \autoref{fig:bb_fit}. The corresponding NIR flux excess is plotted in \autoref{fig:bolometric}. We can see that the continuum shape is well fitted with a 2-component model. 

A NIR excess usually indicates reprocessing of optical photons by newly formed or pre-existing dust. Since we do not expect new dust formation before the optical break, we interpret the first four epochs as emission from pre-existing dust surrounding the progenitor. Dust can form in the dense CSM, and can be heated either radiatively or by shocks, to then emit in the NIR-MIR region depending on the temperature. We discuss this further in \autoref{sec:discussion_nir_excess}.

\begin{figure}
    \centering
    \includegraphics[width=1.\linewidth]{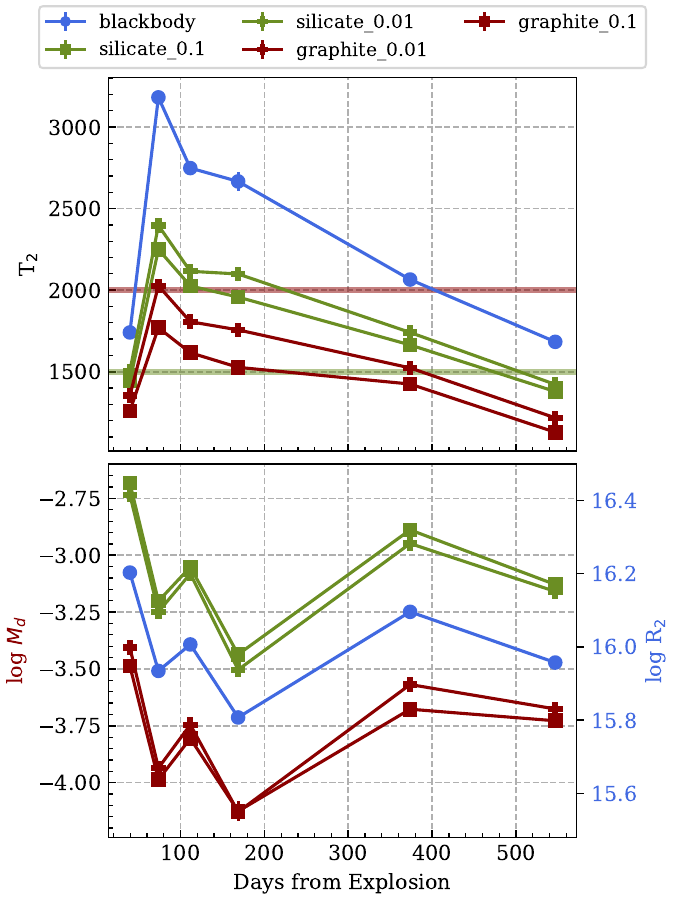}
    \caption{The evolution of dust/blackbody temperature (\textit{top panel}), and the dust-mass/blackbody radius (\textit{bottom panel}) for the multicomponent fit to the spectra. The red and green horizontal lines indicate the sublimation temperature of graphite and silicate grains, respectively.}
    \label{fig:bb_params}
\end{figure}

The time evolution of the fitted parameters for all models is shown in \autoref{fig:bb_params}. The effective temperature of the hot component fades to $\sim 6500$ K by day 112, and remains roughly constant thereafter, consistent with the $(B-V)$ color evolution (\autoref{fig:lk_comparison}). The temperature and the flux evolution of the warm component show an interesting trend. We see an initial rise in temperature and flux of the warm component, which is not usually observed. The NIR flux excess then declines until day $\sim 169$, but we then see a second rise in the NIR flux at day $\sim 374$, hence after the optical break. Since the second rise in the NIR flux occurs coincidentally with the break in optical lightcurves, and the wavelength-dependent extinction in the H lines, we speculate that this rise can be associated with the new dust formation in the post-shock gas and/or the SN ejecta. Therefore, the NIR flux at the last two epochs may contain a contribution from two distinct dust regions.

All four combinations of grain sizes and dust compositions fit the continuum very well, and based on the fit, it is not possible to distinguish among them. The distinctive feature of silicate grains is an emission peak at $\sim 10 \mu$m, which lies well outside the wavelength range considered here. However, from the estimated parameter evolution, we see that the effective temperature of silicate grains is much higher than their sublimation temperature at $T\sim 1500$ K. Silicate grains are generally observed to prefer higher temperatures than graphite grains when fitting the same data \citep{Fox_spitzer_IIn_2011, 2015da_tartaglia}. In this case, temperatures higher than sublimation make it highly unlikely that the dust consists of silicate grains. Additionally, \cite{Fox_spitzer_IIn_2011} studied the NIR-MIR emission from a sample of SNe IIn, and concluded that graphite grains of 0.1 $\mu$m are the preferred composition for pre-existing dust in such SNe.

\section{Results and Discussion} \label{sec:discussion}
We present the comprehensive analysis of the photometric and spectroscopic monitoring campaign of the luminous Type IIn SN~2024kgi. The lightcurves show a slow rise to maximum, lasting around $\sim 41$ days and $\sim 75$ days in the $U$ and $i$ bands, respectively. The lightcurve peaks at $< -19.5$ mag in all optical bands. After the maximum, the lightcurves settle into a smooth decline powered by continued interaction with the CSM. On day $\sim 312$, we note a break in the optical lightcurve, as also observed in SNe~2010jl, 2015da, 2021adxl. This break is likely caused by either a change in the CSM density or the shock sweeping up the dense CSM. 

The bolometric lightcurve follows approximately the same behaviour as the optical lightcurves, with $L \propto t^{-1.0}$ before the break, to steepen to $L \propto t^{-4.5}$ afterward. We model the bolometric lightcurve using a modified version of the CSM interaction model proposed in \cite{CW12} to account for the variable diffusion timescale expected for a moving shock front. The model reveals a very massive CSM envelope of $34.7^{+13.2}_{-7.1} M_{\odot}$ with a density exponent $2.39^{+0.04}_{-0.05}$. The density exponent suggests a deviation from steady wind, with mass-loss rates increasing over time as we approach the SN explosion.

The optical and NIR spectra exhibit prominent H lines throughout the evolution of the SN. The H lines show a narrow component and Lorentzian wings up to day $\sim +80$. Later, the Lorentzian wings are slowly replaced by a combination of intermediate and broad Gaussian components, coming from the CDS and ejecta, respectively. During the same time, broad He lines and \ion{Ca}{2} triplet also start to strengthen. We observe a flux deficit in the red-wing of the $H\alpha$ line as the initial Lorentzian wings fade; however, it is not wavelength dependent at these epochs. Later, after the optical break, the H lines show a wavelength-dependent flux deficit reminiscent of dust formation in the ejecta/CDS.
A few key points are discussed in detail below.

\subsection{Asymmetry}
Interacting SNe often show deviations from spherical symmetry in their CSM structure. In a spherically symmetric CSM, we expect the H lines to be dominated by emission from the CDS near the lightcurve peak. However, when the CSM structure is asymmetric, we should see signatures from the SN ejecta much earlier than expected. In SN 2024kgi, we start to see CDS and ejecta signatures almost simultaneously from day $\sim +80$, as the photosphere recedes from the unshocked CSM. 

Additionally, we expect pre-existing dust in the CSM from the observed NIR excess (\autoref{sec:discussion_nir_excess}). However, the lack of a prominent \ion{Na}{1}D line indicates negligible extinction along the line-of-sight. This can be achieved by an asymmetric dust structure, which in turn implies asymmetry in the CSM where the dust may have formed. A similar scenario is seen in SN~2010jl, which shows signs of preexisting dust with negligible extinction. \cite{2010jl_andrews_csmdust_2011} model the H line profiles with a 3D Monte-Carlo radiative transfer code and find that the dust lies in a torus at a $60^{\rm o} - 80^{\rm o}$ incline to the plane of sky.

Based on the observational signatures discussed here, we expect the CSM around SN~2024kgi to be asymmetric. The exact geometry of the CSM is hard to infer, but structures such as a disk/torus or a bipolar nebula are frequently invoked to explain the observational signatures of SNe~IIn \citep{2015da_smith, asassn14il_dukiya, 2017hcc_smith_andrews}. These types of geometry naturally arise from binary interactions or inhomogeneous winds in LBVs. A recent spectropolarimetric study of SNe~IIn from \cite{IIn_specpol_bilinski} reveals a significant polarization in most events of their sample. The high amount of polarization indicates large ellipticity, which is naturally expected from a disk/torus like or a bipolar CSM structure.

\subsection{Mass loss} \label{sec:discussion_mass_loss}
In \autoref{sec:bolometric}, we derived the following parameters from Model 1: $M_{csm} \approx 34.7\, M_{\odot}$, $s \approx 2.4$, $R_{csm,in} \approx 1.57 \times 10^{14}$ cm, $\rho_{in} \approx 4 \times 10^{-11}$ g~cm$^{-3}$; hence, from the above parameters, we can derive $R_{csm,out}=2.7 \times 10^{16}$ cm.

Therefore, the density structure can be written as

\begin{align*}
\rho =  4.7 \times 10^{-13} \; \left( \frac{r}{10^{15}\, \text{cm}} \right)^{-2.4} & \text{g cm}^{-3}, \\ 
R_{csm,in} & < r < R_{csm,out}. 
\end{align*}

Assuming a wind velocity of $v_w = 130$ km s$^{-1}$, this provides a mass loss rate of 

\begin{align*}
   \dot{M}(r) &= 4\pi r^2 v_w \rho(r) \\
   \dot{M}(r) &= 1.22 \times \left( \frac{r}{10^{15}\, \text{cm}} \right)^{-0.4} M_{\odot}/\text{yr} \\
\end{align*}

Using $r = t_w v_w$, where $t_w$ is the time when CSM was expelled relative to the time of explosion, we get

\begin{align*}
   \dot{M}(t_w) &= 1.74 \times \left( \frac{t_w}{1\, yr} \right)^{-0.4} M_{\odot}/ \text{yr}, \,\,\, 0.38 < t_w < 67.38 \\
\end{align*}

For Model 2, we have: $M_{csm} \approx 2.34\, M_{\odot}$, $s \approx 2.74$, $R_{csm,in} \approx 1.39 \times 10^{14}$ cm, $\rho_{in} \approx 1 \times 10^{-11}$ g~cm$^{-3}$; and therefore $R_{csm,out}=4.9 \times 10^{16}$ cm.
With similar analysis we have,
\begin{align*}
\rho =  4.5 \times 10^{-14} \; \left( \frac{r}{10^{15}\, \text{cm}} \right)^{-2.74} & \text{g cm}^{-3}, \\ 
R_{csm,in} & < r < R_{csm,out}. 
\end{align*}
and
\begin{align*}
   \dot{M}(t_w) &= 0.22 \times \left( \frac{t_w}{1\, yr} \right)^{-0.74} M_{\odot}/ \text{yr}, \,\,\, 0.34 < t_w < 120.51 \\
\end{align*}

For Model 2, the mass-loss rates increased from $\sim 0.01$ $M_{\odot}$/yr in the many decades prior to the explosion to the order of $\sim$ 0.5 $M_{\odot}$/yr in the years leading up to it. The inferred mass-loss rate for Model 1 is roughly an order-of-magnitude higher than Model 1, peaking at $\sim 2 \, M_{\odot}$/yr. Comparable mass-loss rates have been found in the cases of SN~2015da (0.5 -- 0.6 $M_{\odot}$/yr; \citealp{2015da_tartaglia, 2015da_smith}) and SN~2010jl ($\sim 0.8$ $M_{\odot}$/yr; \citealp{2010jl_ofek2014}).
The estimated mass loss rates for both the models are much higher than line-driven wind mass loss of red and yellow supergiants ($10^{-4}-10^{-3}$ M$_{\odot}$ yr$^{-1}$; \citealp{mass_loss_rsg_de_jager, ysg_de_jager, mass_loss_rsg_van_loon, Smith_mass_loss_2014}), or regular LBV winds ($10^{-5}-10^{-3}$ M$_{\odot}$ yr$^{-1}$; \citealp{hillier_2001_eta_carinae, LBV_winds_Vink_deKoter, groh_hillier_lbv_2009, Smith_mass_loss_2014}). 

Giant eruptions of LBVs \citep{smith26,Chugai_1994w_2004, kiewe_iin_sample_2012}, and other explosive mechanisms like pulsational pair instability \citep{Woosley_ppisn_2017, Woosley_smith_1961V_2022} and nuclear burning instabilities \citep{arnett_nuclear_instabilities_2011,sa14} can generate the required mass loss rates. 
However, the eruptive-mass loss episode usually shows signs of multiple ejected shells in photometry and spectra \citep{2009ip_pastorello}, which is at odds with the observed trend of mass-loss rate.
Additionally, the ejected CSM shells from eruptive mass loss episodes are expected to move at much higher velocities as compared to 130 km s\power{-1}\, CSM velocity observed in the case of SN~2024kgi. Another avenue of mass loss in massive stars is binary interactions, which can account for the steady increase in the observed mass-loss rate \citep{Smith_eta_carinae_light_echo_2018, Schorder_binary_IIn_2020} and the asymmetric CSM. And thus, a very promising mechanism for the observed mass-loss history of SN~2024kgi and many other luminous SNe~IIn \citep{2015da_smith, asassn14il_dukiya}. 
Mass-loss rates on the order of $\sim 0.1$--$1 \, M_{\odot}$ would require extreme circumstances, such as unstable mass transfer and common-envelope ejection, in very massive binaries. On the other hand, the Roche-lobe overflow (RLOF) scenario can produce mass loss rates of $10^{-3}$--$10^{-2} M_{\odot}$ \citep{Smith_mass_loss_2014}, which are simply insufficient in this case.

\subsection{Early NIR excess} \label{sec:discussion_nir_excess}
We can study the NIR excess separately before and after the optical break. 
The NIR excess after the optical break likely has contributions from both the newly formed dust in the ejecta/CDS, and residual contribution from the pre-existing dust. Disentangling these components is not trivial with our limited dataset; therefore, we do not discuss it further.

Since there is no wavelength-dependent extinction seen in H lines before the optical break, we can rule out new dust formation (see \autoref{sec:discussion_dust_formation}) at these epochs. Therefore, we interpret the NIR excess prior to the optical break solely due to pre-existing dust in the CSM.
We can calculate the evaporation radius for the dust by the energy balance equation (see Equation 13 of \citealp{Fox_2005ip_2010}):

\begin{equation}
    r_{evap} = \frac{L_{peak}}{16 \sigma T_{SN}^4} \frac{\int B_\lambda(T_{SN}) \kappa(\lambda)d\lambda}{\int B_\lambda(T_{evap}) \kappa(\lambda)d\lambda}.
\end{equation}

\noindent
where $L_{peak}$ is the peak bolometric luminosity of the SN, $T_{SN}$ is the effective temperature of the SN photosphere, $T_{evap}$ is the evaporation temperature of dust ($T\sim 2000\,K$ for graphite dust, $T\sim 1500\,K$ for silicate dust). To calculate the evaporation radius, we consider a wavelength range of $[0.1, 10]\,\mu$m.
From the blackbody fits to the optical photometry, we get a peak temperature $T_{SN} \approx 11,500\,K$, and the peak bolometric luminosity $L_{peak} \approx 3\times 10^{43}$ erg~s$^{-1}$ (see \autoref{sec:bolometric}).
For silicate grains, we estimate an evaporation radius of $\sim 3.3 \times 10^{17}$ cm (or $\sim 130$ light-days). The light travel time to evaporation radius is much higher than the observed delay between the optical and NIR excess peaks (roughly 40 days), again disfavouring a silicate composition for the dust. For graphite grains of 0.01 $\mu$m and 0.1 $\mu$m, we estimate evaporation radii of $2 \times 10^{17}$ cm ($\sim 80$ light-days), and $1 \times 10^{17}$ cm ($\sim 40$ light-days), respectively. The evaporation radius of $\sim 40$ light-days for the 0.1 $\mu$m grains matches well with the delay between the peaks of optical and the NIR lightcurves; however, we note that the NIR lightcurve is not well sampled. The estimated blackbody radius from the NIR excess defines the minimum possible radius of the optically thin dust and, therefore, an evaporation radius much higher than the blackbody radius of $\sim 10^{16}$ cm is physically consistent.

Finally, the evaporation radii allow us to comment on the powering source of the NIR emission. As previously mentioned, pre-existing dust can be heated either radiatively or collisionally by the forward shock. Since we see an NIR excess as soon as day $\sim 40$, ejecta travelling at $\sim 10,000$ km s$^{-1}$ would travel $\sim 3.4\times 10^{15}$ cm by this time. Even ejecta at significantly higher velocities would not reach a distance of $10^{17}$ cm at this epoch. For this reason, direct collisional heating is not a feasible mechanism to power the NIR emission. Therefore, we can conclude that the NIR emission is caused by radiatively heated dust (i.e., an ``IR echo''). 

\subsection{Dust Formation} \label{sec:discussion_dust_formation}
The H line profile of SN~2024kgi shows a flux deficit in the red-wing after day $\sim +80$. A possible explanation is the formation of new dust grains in the post-shock gas or within the ejecta that obscure the receding part of the emitting material. Although dust in ejecta is expected to form after a few hundred days, the formation of CDS in interacting SNe allows for very effective cooling of post-shock gas and can accommodate dust formation even at day $\sim 80$ \citep{Smith_2006jc_dust_2008, Takaya_2006jc_dust_2008, Smith_2005ip_dust}.

Dust formation in the post-shock gas or in the ejecta carries three distinct properties - (i) A wavelength-dependent flux deficit caused by extinction due to dust; (ii) An increased decline rate in the optical lightcurves; (iii) A simultaneous increase in the emitted flux at the NIR-MIR wavelengths. Based on this, we can identify two distinct stages of evolution in the H-line profiles of SN~204kgi. Before the optical break, we see that the flux deficit in the H lines is not wavelength-dependent and the line profiles are similar in H$\beta$, H$\alpha$, and the Pa$\alpha$ lines. As discussed in \autoref{sec:h_lines}, the flux deficit at these epochs can be attributed to occultation from the photosphere or an asymmetry in the CSM distribution, the former being more likely.

At epochs later than the optical break, we observe a wavelength-dependent flux deficit in the H lines, a robust indicator for new dust formation. The rapid fading of the optical continuum may be in part due to newly formed dust, as seen in SN~2010jl. Additionally, we observe an increase in the NIR excess flux after the optical break. Collectively, these observational signatures provide us with a strong case of new dust formation in the post-shock gas and/or the ejecta. Similarly, in the case of SN~2010jl, new dust formation is expected only after the optical break \citep{2010jl_Sarangi_2018}. A remarkable coincidence is the alignment of dust formation timing with the break in the forward shock for both SNe~2010jl and 2024kgi. It may be a chance coincidence, but it might also indicate a causal relationship. We speculate that the termination of the constant energy input from the forward shock finally allows the post-shock gas to cool rapidly, commencing dust formation.

\section{Conclusions} \label{sec:conclusions}
The photometric and spectroscopic evolution of SN~2024kgi reveals the hallmark signatures of strong ejecta-CSM interaction. Early spectra are characterized by narrow H and He lines with Lorentzian wings. After day $\sim +80$, the line profiles evolve into a combination of intermediate (FWHM 1500--2000 km s$^{-1}$) and broad (FWHM 6000--10000 km s$^{-1}$) components accompanied by red-wing flux deficit in H lines. The lightcurves exhibit a broken power law decline, radiating $\sim 2 \times 10^{50}$ ergs over the observed period. By modeling the post-peak bolometric lightcurve, we determine the CSM mass to be $2.34^{+11.8}_{-1.6} M_{\odot}$ with a density profile $\rho_{csm} \propto r^{-2.74}$.
Post the lightcurve break, wavelength dependence is seen in the red-wing flux deficit of H lines. 
A NIR excess is detected as early as day $\sim +40$, and is well reproduced by a combination of blackbody and optically thin dust emission.

These observations are broadly consistent with a SN exploding within a dense, asymmetric disk/torus like CSM produced by binary-driven mass loss. The inferred CSM density profile suggests an increasing mass-loss rate toward core collapse expected from binary interactions.
Preexisting dust inside this CSM envelope produces the NIR excess observed at early times. The asymmetric CSM allows for the CDS and the freely expanding ejecta to become visible to the observer shortly after the peak luminosity. 
The observed light-curve break at around day 300 is consistent with the forward shock reaching the outer edge of the dense CSM, after which dust formation in the cooling ejecta and post-shock gas accounts for the enhanced NIR emission and the wavelength-dependent attenuation of the H line profiles.

Overall, SN~2024kgi supports a picture in which at least some luminous, slowly evolving SNe IIn arise from prolonged, asymmetric mass loss in interacting massive binaries rather than isolated single-star progenitors. Expanding such studies to a larger sample of events will be essential for establishing the relative importance of binary evolution, eruptive mass loss, and other progenitor channels in shaping the diversity of the SNe IIn population.

Although the proposed scenario provides a self-consistent explanation for the observations, the inferred mass loss rates of the order of $\sim 0.1 M_{\odot}$/yr are difficult to reconcile with the ordinary binary RLOF scheme. Such extreme values would likely require unstable mass transfer and common-envelope ejection in very massive binary systems \citep{Smith_mass_loss_2014}. On the other hand, eruptive mechanisms such as giant eruptions of LBVs and pulsational pair instabilities may account for the observed mass-loss rates. However, the relatively slow outflow velocity of the CSM and steadily increasing mass loss is more naturally explained by prolonged binary interaction rather than a single explosive mass loss episode. 
Additionally, the available data do not allow us to unambiguously determine the origin of the late-time NIR excess. Future high-quality NIR--MIR observations from facilities such as JWST, complemented by wide-field surveys from Euclid, will enable more robust constraints on dust formation channels and composition. At the same time, Nancy Grace Roman Telescope will discover and monitor substantially larger samples of SNe IIn, their precursor and possibly their progenitors, with its wide-field, Hubble-class imaging capabilities. Together, these facilities will provide the observational leverage needed to connect progenitor channels, mass loss histories, and dust formation across the SNe IIn population.

\software{Astropy \citep{astropy:2013, astropy:2018, astropy:2022}, Scipy \citep{2020SciPy-NMeth}, emcee \citep{emcee_Foreman-Mackey_2013}}, IRAF \citep{iraf_tody}, PSFEx \citep{psfex_ascl}.

\facility{NED}

\section*{Acknowledgements}
We thank the anonymous referee for providing us with valuable suggestions and scientific insights that enhanced the quality of the paper.
This work uses data from the Las Cumbres Observatory Global Telescope network. The LCO group is supported by U.S. National Science Foundation (NSF) grants AST-2308113 and AST-1911151. This research has made use of the NASA/IPAC Extragalactic Database, which is funded by the National Aeronautics and Space Administration and operated by the California Institute of Technology. Some observations reported here were obtained at the MMT Observatory, a joint facility of the University of Arizona and the Smithsonian Institution. We acknowledge Weizmann Interactive Supernova data REPository \url{http://wiserep.weizmann.ac.il} (WISeREP, \citealp{yaron_wiserep}). 

ND and KM acknowledge the support from the BRICS grant DST/ICD/BRICS/Call-5/CoNMuTraMO/2023 (G) funded by the Department of Science and Technology (DST), India. MD acknowledges the Innovation in Science Pursuit for Inspired Research (INSPIRE) fellowship award (DST/INSPIRE Fellowship/2020/IF200251) for this work. NF acknowledges support from the National Science Foundation Graduate Research Fellowship Program under Grant No. DGE-2137419. AP, AR, LT and GV acknowledge support from the PRIN-INAF 2022, “Shedding light on the nature of gap transients: from the observations to the models”. 


\bibliography{references}
\bibliographystyle{aasjournal}

\appendix
\counterwithin{figure}{section}
\counterwithin{table}{section}

\section{Chevalier Solutions} \label{sec:chevalier_solutions}
We provide a brief summary of self-similar hydrodynamical solution obtained by \cite{Chevalier_1982_self-similar} for a case of ejecta with a power-law density profile ($\rho_{ej} \propto r^{-n}$) interacting with a CSM with power-law density profile ($\rho_{csm} \propto r^{-s}$). For the ejecta, we assume a density profile of $\rho_{ej} = g_n t^{n-3} r^{-n}$, where $g_n = 1/(4\pi(\delta-n))[2(5-\delta)(n-5)E_{sn}]^{(n-3)/2} / [(3-\delta)(n-3)M_{ej}]^{(n-5)/2}$ is a scaling parameter, $n$ is the power-law index of outer ejecta ($n > 5$), $\delta$ is the power-law index of inner ejecta, $M_{ej}$ is the mass of the ejecta, and $E_{sn}$ is the kinetic energy of the ejecta. Furthermore, we assume a CSM density profile of $\rho_{csm} = q r^{-s}$, where $s$ is the power-law index of CSM and $q$ is the CSM mass loading parameter.

The radii of forward and reverse shocks are given by

\begin{equation}
    R_{fs}(t) = R_{csm,in} + \beta_{fs} \left(\frac{Ag_n}{q}\right)^{\frac{1}{n-s}} t^{-m}
\end{equation}

\begin{equation}
    R_{rs}(t) = R_{csm,in} + \beta_{rs} \left(\frac{Ag_n}{q}\right)^{\frac{1}{n-s}} t^{-m}
\end{equation}

\noindent
where $m = (n-3)/(n-s)$, $R_{csm,in}$ is the inner radius of the CSM; $A, \beta_{fs}$, and $\beta_{rs}$ are constants of $n$ and $s$. \cite{Chevalier_1982_self-similar} tabulated these constants for a handful of combinations of $n$ \& $s$. Later, \cite{Jiang_IIn_analytical_mosfit} provided a numerical method to find these constants for any given value of $n$ \& $s$.

Once these radii are known, the velocity $v_{sh}$ and acceleration $a_{sh}$ can be calculated. The mass swept up ($M_{sw}$) by the forward shock as a function of time can be obtained by integrating between $R_{csm,in}$ and $R_{fs}$. Similarly, the mass swept up by the reverse shock can be found by integrating between $R_{rs}$ and $R_{sn}$, where $R_{sn}$ is the radius of the SN photosphere.

In this case, the luminosity

\begin{equation}
\begin{split}
L = \epsilon\frac{d E}{dt} = & \epsilon\frac{d}{dt}\left(\frac{1}{2}M_{sw}v_{sh}^2\right) \\
  = & \epsilon\left(M_{sw}v_{sh}a_{sh} + \frac{1}{2}\dot{M}_{sw}v_{sh} ^2\right)
\end{split}
\end{equation}

adapted for the forward and reverse shocks becomes respectively:

\begin{equation}
    L_{fs}(t) =\left\{
\begin{array}{ll}
    \epsilon_{fs} C_{fs} \, t^{\alpha}, & \text{if } 0 < t < t_{fs} \\
    0, & \text{otherwise}
\end{array}
\right.
\end{equation}

\begin{equation}
    L_{rs}(t) =\left\{
\begin{array}{ll}
    \epsilon_{rs} C_{rs} \, t^{\alpha}, & \text{if } 0 < t < t_{rs} \\
    0, & \text{otherwise}
\end{array}
\right.
\end{equation}

\noindent
Here, $\alpha = (2n+6s-ns-15)/(n-s)$, $\epsilon$ is the efficiency factor for the conversion of kinetic energy into radiation; $C_{fs}$ is a constant of $n, s, g_n, q, \beta_{fs}, \text{ and } A$; $C_{rs}$ is a constant of $n, s, g_n, q, \beta_{rs}, \text{ and } A$; $t_{fs}$ and $t_{rs}$ denote the termination timescales for the forward and reverse shocks, respectively.

\section{Observation log}
\startlongtable
\begin{deluxetable*}{ccccccccc}
\tabletypesize{\footnotesize}
\tablecaption{Log of photometric observations of SN~2024kgi.} 
\label{tab:photometry_log}
\tablehead{
\colhead{Date} & \colhead{Phase\tablenotemark{$\dagger$}} & \colhead{U} & \colhead{B} & \colhead{V} & \colhead{g} & \colhead{r} & \colhead{i} & \colhead{telescope}\\
\colhead{(yyyy-mm-dd)} & \colhead{(day)} & \colhead{(mag)} & \colhead{(mag)} & \colhead{(mag)} & \colhead{(mag)} & \colhead{(mag)} & \colhead{(mag)} & \colhead{}
}
\startdata
2024-06-14 & 15.82 & 16.53$\pm$0.03 & 17.23$\pm$0.01 & 17.17$\pm$0.02 & 17.05$\pm$0.01 & 17.03$\pm$0.01 & 17.24$\pm$0.02 & LCO 1m0-13 \\
2024-06-17 & 19.04 & 16.26$\pm$0.02 & 16.99$\pm$0.01 & 16.93$\pm$0.01 & 16.85$\pm$0.01 & 16.83$\pm$0.01 & 17.04$\pm$0.01 & LCO 1m0-05 \\
2024-06-21 & 22.75 & 16.15$\pm$0.05 & 16.95$\pm$0.03 & 16.82$\pm$0.03 & 16.72$\pm$0.02 & 16.65$\pm$0.02 & 16.85$\pm$0.03 & LCO 1m0-10 \\
2024-06-24 & 26.48 & 16.01$\pm$0.02 & 16.84$\pm$0.01 & 16.66$\pm$0.01 & 16.61$\pm$0.01 & 16.57$\pm$0.01 & 16.67$\pm$0.02 & LCO 1m0-03 \\
2024-06-28 & 30.36 & 16.11$\pm$0.02 & 16.82$\pm$0.01 & 16.60$\pm$0.01 & 16.60$\pm$0.01 & 16.54$\pm$0.01 & 16.61$\pm$0.02 & LCO 1m0-03 \\
2024-07-02 & 34.07 & 16.08$\pm$0.02 & 16.79$\pm$0.01 & 16.61$\pm$0.01 & 16.61$\pm$0.01 & 16.53$\pm$0.01 & 16.62$\pm$0.02 & LCO 1m0-04 \\
2024-07-06 & 37.78 & 16.13$\pm$0.02 & 16.75$\pm$0.01 & 16.52$\pm$0.01 & 16.53$\pm$0.01 & 16.44$\pm$0.01 & 16.50$\pm$0.01 & LCO 1m0-01 \\
2024-07-10 & 41.84 & 16.19$\pm$0.02 & 16.74$\pm$0.01 & 16.52$\pm$0.01 & 16.53$\pm$0.01 & 16.40$\pm$0.01 & 16.47$\pm$0.01 & LCO 1m0-01 \\
2024-07-14 & 45.93 & 16.27$\pm$0.02 & 16.83$\pm$0.01 & 16.54$\pm$0.02 & 16.60$\pm$0.01 & 16.40$\pm$0.01 & 16.44$\pm$0.01 & LCO 1m0-04 \\
2024-07-18 & 49.92 & -- & -- & -- & 16.64$\pm$0.01 & -- & -- & LCO 1m0-04 \\
2024-07-19 & 51.45 & 16.38$\pm$0.06 & 16.86$\pm$0.02 & 16.49$\pm$0.02 & 16.55$\pm$0.01 & 16.36$\pm$0.01 & 16.39$\pm$0.02 & LCO 1m0-11 \\
2024-07-23 & 55.29 & 16.44$\pm$0.05 & 16.88$\pm$0.02 & 16.52$\pm$0.02 & 16.61$\pm$0.01 & 16.39$\pm$0.01 & 16.40$\pm$0.02 & LCO 1m0-11 \\
2024-07-27 & 59.00 & 16.44$\pm$0.02 & 16.84$\pm$0.01 & 16.51$\pm$0.01 & 16.61$\pm$0.01 & 16.37$\pm$0.01 & 16.38$\pm$0.01 & LCO 1m0-05 \\
2024-08-01 & 63.76 & 16.59$\pm$0.03 & 16.94$\pm$0.02 & 16.53$\pm$0.01 & 16.69$\pm$0.01 & 16.41$\pm$0.01 & 16.39$\pm$0.01 & LCO 1m0-10 \\
2024-08-05 & 67.89 & 16.65$\pm$0.03 & 16.93$\pm$0.01 & 16.52$\pm$0.02 & 16.65$\pm$0.01 & 16.41$\pm$0.01 & 16.37$\pm$0.01 & LCO 1m0-01 \\
2024-08-10 & 73.43 & 16.70$\pm$0.04 & 17.04$\pm$0.01 & 16.55$\pm$0.01 & 16.69$\pm$0.01 & 16.40$\pm$0.01 & 16.35$\pm$0.01 & LCO 1m0-11 \\
2024-08-14 & 77.59 & 16.80$\pm$0.05 & 17.07$\pm$0.02 & 16.54$\pm$0.02 & 16.74$\pm$0.01 & 16.44$\pm$0.02 & 16.36$\pm$0.02 & LCO 1m0-10 \\
2024-08-14 & 77.61 & 16.86$\pm$0.04 & 17.04$\pm$0.02 & 16.60$\pm$0.02 & 16.70$\pm$0.01 & 16.39$\pm$0.01 & 16.36$\pm$0.01 & LCO 1m0-13 \\
2024-08-18 & 81.15 & 16.77$\pm$0.02 & 17.08$\pm$0.01 & 16.58$\pm$0.01 & 16.77$\pm$0.01 & 16.42$\pm$0.01 & 16.34$\pm$0.01 & LCO 1m0-08 \\
2024-08-21 & 84.61 & 16.92$\pm$0.03 & 17.10$\pm$0.02 & 16.60$\pm$0.02 & 16.78$\pm$0.01 & 16.40$\pm$0.01 & 16.35$\pm$0.02 & LCO 1m0-01 \\
2024-08-21 & 83.74 & 16.89$\pm$0.02 & 17.09$\pm$0.07 & 16.59$\pm$0.05 & 16.84$\pm$0.03 & 16.42$\pm$0.02 & 16.35$\pm$0.02 & LCO 1m0-12 \\
2024-08-21 & 83.73 & -- & -- & -- & 16.75$\pm$0.04 & 16.39$\pm$0.03 & 16.37$\pm$0.03 & LCO 1m0-13 \\
2024-08-25 & 87.78 & 16.93$\pm$0.03 & 17.17$\pm$0.01 & 16.66$\pm$0.01 & 16.84$\pm$0.01 & 16.45$\pm$0.01 & 16.38$\pm$0.04 & LCO 1m0-01 \\
2024-08-25 & 87.82 & 16.92$\pm$0.01 & 17.16$\pm$0.02 & 16.62$\pm$0.02 & 16.83$\pm$0.01 & 16.43$\pm$0.01 & 16.38$\pm$0.02 & LCO 1m0-05 \\
2024-08-25 & 87.83 & 16.96$\pm$0.02 & 17.12$\pm$0.01 & 16.60$\pm$0.01 & 16.84$\pm$0.01 & 16.45$\pm$0.01 & 16.40$\pm$0.01 & LCO 1m0-14 \\
2024-08-29 & 92.31 & 17.02$\pm$0.03 & 17.25$\pm$0.02 & 16.69$\pm$0.01 & 16.89$\pm$0.01 & 16.47$\pm$0.01 & 16.40$\pm$0.01 & LCO 1m0-11 \\
2024-09-01 & 94.70 & -- & 17.24$\pm$0.09 & 16.69$\pm$0.05 & 16.99$\pm$0.15 & 16.51$\pm$0.07 & 16.48$\pm$0.11 & Asiago Schmidt 67/92 \\
2024-09-01 & 95.66 & 17.06$\pm$0.02 & 17.31$\pm$0.01 & 16.73$\pm$0.02 & 16.94$\pm$0.01 & 16.53$\pm$0.01 & 16.45$\pm$0.01 & LCO 1m0-14 \\
2024-09-05 & 99.37 & 17.06$\pm$0.03 & 17.36$\pm$0.01 & 16.77$\pm$0.01 & 17.00$\pm$0.01 & 16.51$\pm$0.01 & 16.43$\pm$0.01 & LCO 1m0-11 \\
2024-09-09 & 103.08 & 17.20$\pm$0.02 & 17.37$\pm$0.01 & 16.80$\pm$0.01 & 17.01$\pm$0.01 & 16.57$\pm$0.01 & 16.47$\pm$0.01 & LCO 1m0-08 \\
2024-09-10 & 103.70 & -- & -- & -- & 17.08$\pm$0.11 & -- & -- & Asiago Schmidt 67/92 \\
2024-09-10 & 104.29 & 17.13$\pm$0.03 & 17.41$\pm$0.01 & 16.81$\pm$0.01 & 17.03$\pm$0.01 & 16.57$\pm$0.01 & 16.48$\pm$0.01 & LCO 1m0-11 \\
2024-09-13 & 107.62 & 17.38$\pm$0.05 & 17.46$\pm$0.02 & 16.85$\pm$0.02 & 17.09$\pm$0.01 & 16.59$\pm$0.02 & 16.54$\pm$0.01 & LCO 1m0-10 \\
2024-09-17 & 111.65 & 17.32$\pm$0.04 & 17.47$\pm$0.02 & 16.87$\pm$0.02 & 17.11$\pm$0.01 & 16.61$\pm$0.02 & 16.59$\pm$0.02 & LCO 1m0-14 \\
2024-09-21 & 114.70 & -- & 17.56$\pm$0.07 & 16.92$\pm$0.05 & -- & -- & 16.60$\pm$0.11 & Asiago Schmidt 67/92 \\
2024-09-21 & 115.58 & 17.41$\pm$0.04 & 17.53$\pm$0.02 & 16.94$\pm$0.01 & 17.16$\pm$0.01 & 16.64$\pm$0.01 & 16.56$\pm$0.01 & LCO 1m0-10 \\
2024-09-30 & 124.24 & 17.59$\pm$0.03 & 17.65$\pm$0.01 & 17.02$\pm$0.01 & 17.27$\pm$0.01 & 16.73$\pm$0.01 & 16.66$\pm$0.01 & LCO 1m0-11 \\
2024-10-03 & 127.62 & 17.60$\pm$0.02 & 17.68$\pm$0.01 & 17.07$\pm$0.01 & 17.31$\pm$0.01 & 16.80$\pm$0.01 & 16.72$\pm$0.01 & LCO 1m0-01 \\
2024-10-07 & 131.56 & 17.69$\pm$0.04 & 17.73$\pm$0.01 & 17.11$\pm$0.02 & 17.41$\pm$0.01 & 16.80$\pm$0.01 & 16.77$\pm$0.02 & LCO 1m0-10 \\
2024-10-12 & 136.22 & 17.80$\pm$0.09 & 17.79$\pm$0.03 & 17.14$\pm$0.02 & 17.38$\pm$0.01 & 16.80$\pm$0.01 & 16.78$\pm$0.02 & LCO 1m0-03 \\
2024-10-14 & 137.70 & -- & 17.79$\pm$0.11 & 17.13$\pm$0.12 & 17.46$\pm$0.10 & 16.95$\pm$0.10 & 16.74$\pm$0.17 & Asiago Schmidt 67/92 \\
2024-10-16 & 140.04 & 17.75$\pm$0.08 & 17.84$\pm$0.04 & 17.16$\pm$0.03 & 17.43$\pm$0.02 & 16.85$\pm$0.02 & 16.81$\pm$0.03 & LCO 1m0-08 \\
2024-10-19 & 143.66 & 17.79$\pm$0.03 & 17.83$\pm$0.01 & 17.17$\pm$0.01 & 17.45$\pm$0.01 & 16.85$\pm$0.01 & 16.84$\pm$0.01 & LCO 1m0-14 \\
2024-10-24 & 148.24 & 17.88$\pm$0.04 & 17.89$\pm$0.01 & 17.22$\pm$0.01 & 17.48$\pm$0.01 & 16.89$\pm$0.01 & 16.89$\pm$0.01 & LCO 1m0-03 \\
2024-10-28 & 151.81 & 17.90$\pm$0.03 & 17.91$\pm$0.01 & 17.27$\pm$0.01 & 17.52$\pm$0.01 & 16.92$\pm$0.01 & 16.95$\pm$0.01 & LCO 1m0-09 \\
2024-10-31 & 155.61 & -- & -- & -- & 17.48$\pm$0.12 & -- & 16.95$\pm$0.07 & Copernico 1.82m \\
2024-10-31 & 155.56 & 17.92$\pm$0.05 & 17.97$\pm$0.01 & 17.33$\pm$0.01 & 17.55$\pm$0.01 & 16.93$\pm$0.01 & 16.97$\pm$0.01 & LCO 1m0-10 \\
2024-11-05 & 159.97 & -- & 18.01$\pm$0.02 & -- & 17.56$\pm$0.01 & -- & -- & LCO 1m0-06 \\
2024-11-05 & 160.12 & 17.90$\pm$0.04 & 18.02$\pm$0.01 & 17.36$\pm$0.01 & 17.59$\pm$0.01 & 16.97$\pm$0.01 & 16.98$\pm$0.01 & LCO 1m0-11 \\
2024-11-10 & 165.19 & 17.91$\pm$0.11 & 17.98$\pm$0.04 & 17.36$\pm$0.03 & 17.63$\pm$0.04 & 17.01$\pm$0.02 & 17.09$\pm$0.04 & LCO 1m0-11 \\
2024-11-11 & 166.60 & 18.03$\pm$0.05 & 18.12$\pm$0.03 & 17.38$\pm$0.02 & 17.64$\pm$0.02 & 17.04$\pm$0.02 & 17.04$\pm$0.02 & LCO 1m0-01 \\
2024-11-11 & 165.80 & 17.95$\pm$0.05 & 18.03$\pm$0.03 & 17.40$\pm$0.02 & 17.59$\pm$0.01 & 16.98$\pm$0.01 & 17.04$\pm$0.02 & LCO 1m0-05 \\
2024-11-15 & 170.52 & 17.94$\pm$0.08 & 18.06$\pm$0.03 & 17.41$\pm$0.02 & 17.68$\pm$0.02 & -- & 17.08$\pm$0.01 & LCO 1m0-10 \\
2024-11-17 & 172.52 & 18.09$\pm$0.03 & 18.07$\pm$0.01 & 17.42$\pm$0.01 & 17.69$\pm$0.01 & 17.02$\pm$0.01 & 17.08$\pm$0.01 & LCO 1m0-12 \\
2024-11-22 & 176.90 & 18.05$\pm$0.03 & 18.10$\pm$0.01 & 17.45$\pm$0.01 & 17.69$\pm$0.01 & 17.04$\pm$0.01 & 17.11$\pm$0.01 & LCO 1m0-08 \\
2024-11-26 & 180.76 & 17.93$\pm$0.04 & 18.10$\pm$0.01 & 17.48$\pm$0.02 & 17.70$\pm$0.01 & 17.02$\pm$0.01 & 17.15$\pm$0.01 & LCO 1m0-05 \\
2024-11-30 & 184.75 & 18.12$\pm$0.04 & 18.14$\pm$0.01 & 17.48$\pm$0.01 & 17.75$\pm$0.01 & 17.07$\pm$0.01 & 17.19$\pm$0.01 & LCO 1m0-09 \\
2024-12-04 & 188.73 & 18.15$\pm$0.06 & 18.10$\pm$0.02 & 17.48$\pm$0.02 & 17.72$\pm$0.01 & 17.03$\pm$0.01 & 17.18$\pm$0.01 & LCO 1m0-05 \\
2024-12-09 & 194.61 & 18.07$\pm$0.04 & 18.15$\pm$0.02 & 17.49$\pm$0.06 & -- & -- & -- & LCO 1m0-14 \\
2024-12-11 & 195.87 & -- & 18.18$\pm$0.05 & 17.53$\pm$0.04 & 17.66$\pm$0.03 & 16.99$\pm$0.02 & 17.18$\pm$0.02 & LCO 1m0-06 \\
2024-12-15 & 199.84 & 18.22$\pm$0.05 & 18.19$\pm$0.02 & 17.54$\pm$0.02 & 17.79$\pm$0.02 & 17.05$\pm$0.01 & 17.26$\pm$0.02 & LCO 1m0-08 \\
2024-12-18 & 203.57 & 18.22$\pm$0.04 & 18.19$\pm$0.02 & 17.57$\pm$0.02 & 17.82$\pm$0.01 & 17.10$\pm$0.01 & 17.27$\pm$0.01 & LCO 1m0-01 \\
2024-12-23 & 207.84 & 18.22$\pm$0.04 & 18.21$\pm$0.01 & 17.56$\pm$0.01 & 17.76$\pm$0.01 & 17.07$\pm$0.01 & 17.27$\pm$0.02 & LCO 1m0-08 \\
2025-01-02 & 218.57 & 18.25$\pm$0.03 & 18.28$\pm$0.02 & 17.62$\pm$0.02 & 17.86$\pm$0.01 & 17.13$\pm$0.01 & 17.34$\pm$0.02 & LCO 1m0-14 \\
2025-01-11 & 227.55 & 18.30$\pm$0.07 & 18.31$\pm$0.03 & 17.65$\pm$0.02 & 17.90$\pm$0.02 & 17.14$\pm$0.01 & 17.36$\pm$0.02 & LCO 1m0-14 \\
2025-01-17 & 233.53 & 18.35$\pm$0.05 & 18.36$\pm$0.02 & 17.71$\pm$0.02 & 17.97$\pm$0.01 & 17.17$\pm$0.01 & 17.43$\pm$0.01 & LCO 1m0-01 \\
2025-05-06 & 342.49 & -- & -- & -- & 18.75$\pm$0.01 & 17.72$\pm$0.01 & 18.33$\pm$0.03 & LCO 1m0-11 \\
2025-05-07 & 342.85 & 19.15$\pm$0.11 & 19.09$\pm$0.03 & 18.46$\pm$0.02 & -- & -- & -- & LCO 1m0-13 \\
2025-05-09 & 345.13 & 19.53$\pm$0.14 & 19.26$\pm$0.03 & 18.51$\pm$0.02 & 18.79$\pm$0.01 & 17.71$\pm$0.01 & 18.30$\pm$0.02 & LCO 1m0-08 \\
2025-05-15 & 351.12 & 19.26$\pm$0.14 & 19.28$\pm$0.05 & 18.57$\pm$0.06 & 18.84$\pm$0.03 & 17.76$\pm$0.02 & 18.44$\pm$0.04 & LCO 1m0-05 \\
2025-05-21 & 357.06 & 19.55$\pm$0.15 & 19.35$\pm$0.08 & -- & -- & -- & -- & LCO 1m0-05 \\
2025-05-23 & 359.05 & 19.61$\pm$0.16 & 19.38$\pm$0.03 & 18.71$\pm$0.03 & 18.96$\pm$0.02 & 17.87$\pm$0.01 & 18.46$\pm$0.03 & LCO 1m0-04 \\
2025-05-29 & 365.08 & 19.67$\pm$0.08 & 19.44$\pm$0.02 & 18.75$\pm$0.02 & 18.99$\pm$0.01 & 17.89$\pm$0.01 & 18.58$\pm$0.03 & LCO 1m0-05 \\
2025-06-04 & 371.12 & 19.47$\pm$0.31 & 19.50$\pm$0.04 & 18.81$\pm$0.03 & 19.06$\pm$0.02 & 17.95$\pm$0.01 & 18.56$\pm$0.03 & LCO 1m0-06 \\
2025-06-10 & 377.11 & 19.57$\pm$0.20 & 19.62$\pm$0.06 & 18.92$\pm$0.03 & 19.17$\pm$0.02 & 18.09$\pm$0.01 & 18.67$\pm$0.02 & LCO 1m0-09 \\
2025-06-16 & 382.83 & 19.50$\pm$0.26 & 19.63$\pm$0.11 & 18.95$\pm$0.09 & 19.29$\pm$0.07 & 18.12$\pm$0.04 & 18.55$\pm$0.09 & LCO 1m0-12 \\
2025-06-22 & 389.14 & -- & 19.71$\pm$0.04 & 19.11$\pm$0.03 & 19.32$\pm$0.02 & 18.14$\pm$0.01 & 18.86$\pm$0.04 & LCO 1m0-06 \\
2025-06-24 & 390.96 & 20.00$\pm$0.12 & 19.84$\pm$0.03 & 19.11$\pm$0.03 & 19.34$\pm$0.02 & 18.22$\pm$0.02 & 18.86$\pm$0.03 & LCO 1m0-05 \\
2025-07-03 & 400.01 & 19.92$\pm$0.10 & 19.91$\pm$0.03 & 19.26$\pm$0.03 & 19.52$\pm$0.02 & 18.36$\pm$0.02 & 19.00$\pm$0.03 & LCO 1m0-09 \\
2025-07-11 & 408.02 & -- & 19.85$\pm$0.11 & 19.39$\pm$0.07 & 19.60$\pm$0.06 & 18.45$\pm$0.03 & 19.05$\pm$0.05 & LCO 1m0-09 \\
2025-07-11 & 407.83 & -- & 19.65$\pm$0.18 & 19.46$\pm$0.24 & 19.57$\pm$0.26 & -- & -- & LCO 1m0-13 \\
2025-07-19 & 416.67 & -- & 20.02$\pm$0.24 & -- & -- & -- & -- & LCO 1m0-13 \\
2025-07-20 & 416.89 & 20.11$\pm$0.19 & 20.15$\pm$0.04 & 19.44$\pm$0.03 & 19.73$\pm$0.02 & 18.53$\pm$0.02 & 19.18$\pm$0.03 & LCO 1m0-04 \\
2025-07-29 & 425.76 & 20.32$\pm$0.27 & 20.28$\pm$0.06 & 19.52$\pm$0.05 & 19.84$\pm$0.03 & 18.62$\pm$0.02 & 19.20$\pm$0.06 & LCO 1m0-12 \\
2025-08-06 & 434.60 & -- & -- & -- & 20.01$\pm$0.14 & 18.79$\pm$0.05 & -- & LCO 1m0-12 \\
2025-08-16 & 444.04 & 20.07$\pm$0.16 & 20.35$\pm$0.08 & 19.44$\pm$0.07 & 19.99$\pm$0.04 & 18.87$\pm$0.03 & -- & LCO 1m0-05 \\
2025-08-24 & 452.31 & 20.24$\pm$0.26 & 20.56$\pm$0.06 & 19.87$\pm$0.05 & 20.15$\pm$0.04 & 19.01$\pm$0.02 & 19.67$\pm$0.06 & LCO 1m0-11 \\
2025-09-02 & 461.06 & -- & 20.76$\pm$0.08 & 20.12$\pm$0.05 & 20.34$\pm$0.04 & 19.11$\pm$0.02 & 19.75$\pm$0.05 & LCO 1m0-06 \\
2025-09-11 & 469.78 & 20.52$\pm$0.18 & 20.67$\pm$0.07 & 20.07$\pm$0.07 & 20.26$\pm$0.05 & -- & 19.88$\pm$0.10 & LCO 1m0-05 \\
2025-09-13 & 471.86 & 20.68$\pm$0.31 & 20.78$\pm$0.12 & 20.14$\pm$0.09 & 20.33$\pm$0.07 & 19.32$\pm$0.04 & 20.01$\pm$0.10 & LCO 1m0-01 \\
2025-09-22 & 480.87 & -- & 20.62$\pm$0.08 & 20.09$\pm$0.08 & 20.38$\pm$0.05 & 19.17$\pm$0.04 & 19.79$\pm$0.10 & LCO 1m0-04 \\
2025-09-30 & 489.50 & -- & 20.68$\pm$0.14 & 20.40$\pm$0.15 & 20.41$\pm$0.08 & 19.54$\pm$0.04 & 20.05$\pm$0.12 & LCO 1m0-12 \\
2025-10-18 & 507.50 & 21.25$\pm$0.45 & 21.16$\pm$0.12 & 20.44$\pm$0.11 & 20.69$\pm$0.08 & 19.65$\pm$0.04 & -- & LCO 1m0-12 \\
2025-10-29 & 517.80 & 20.77$\pm$0.27 & 21.24$\pm$0.13 & 20.58$\pm$0.09 & 20.86$\pm$0.07 & 19.84$\pm$0.05 & 20.34$\pm$0.10 & LCO 1m0-09 \\
2025-11-09 & 528.81 & 21.25$\pm$0.28 & 21.15$\pm$0.08 & 20.57$\pm$0.07 & 20.90$\pm$0.06 & 19.83$\pm$0.04 & 20.40$\pm$0.12 & LCO 1m0-09 \\
2025-11-20 & 539.76 & -- & 21.40$\pm$0.08 & 20.69$\pm$0.08 & 20.88$\pm$0.05 & 20.06$\pm$0.04 & 20.62$\pm$0.12 & LCO 1m0-09 \\
2025-12-03 & 552.84 & -- & -- & 20.69$\pm$0.26 & 20.98$\pm$0.28 & -- & -- & LCO 1m0-06 \\
2025-12-14 & 563.85 & 21.17$\pm$0.30 & -- & -- & -- & 20.06$\pm$0.07 & -- & LCO 1m0-06 \\
2025-12-25 & 574.83 & 21.72$\pm$0.49 & 21.71$\pm$0.19 & 20.92$\pm$0.12 & -- & 20.29$\pm$0.08 & -- & LCO 1m0-06 \\
\enddata
\tablenotetext{\dagger}{The phase is given with respect to the explosion epoch of JD 2460459.8.}
\end{deluxetable*}
\newpage
\begin{table*}
\centering
\caption{Log of spectroscopic observations of SN~2024kgi. The resolution, if measured from the \ion{O}{3} 6300\AA\, sky line whenever possible. For the spectra marked with (*), the quoted number reflects the typical resolution expected from the instrumental setup.}
\label{tab:spectra_log}
\begin{tabular}{ccccccc}
\hline\hline
Date & Phase\tablenotemark{$\dagger$} & Telescope $+$ Instrument & Grism & Wavelength Range & Slit width & Resolution \\
(yy-mm-dd) & (day) & & & (\AA) & (arcsec) & ($\lambda / \Delta\lambda$) \\\hline
2024-06-12 & 14.20 & LCO+FLOYDS & red/blu & 3200--10000 & 2.0 & 373 \\
2024-06-21 & 23.29 & LCO+FLOYDS & red/blu & 3200--10000 & 2.0 & 450* \\
2024-06-29 & 31.14 & LCO+FLOYDS & red/blu & 3200--10000 & 2.0 & 401 \\
2024-07-07 & 39.20 & LCO+FLOYDS & red/blu & 3200--10000 & 2.0 & 406 \\
2024-07-09 & 41.28 & IRTF+SpeX & ShortXD & 7000--25500 & 0.8 & 800* \\
2024-07-13 & 45.21 & LCO+FLOYDS & red/blu & 3200--10000 & 2.0 & 409 \\
2024-07-21 & 53.12 & LCO+FLOYDS & red/blu & 3200--10000 & 2.0 & 509 \\
2024-08-01 & 64.13 & LCO+FLOYDS & red/blu & 3200--10000 & 2.0 & 420 \\
2024-08-09 & 72.13 & LCO+FLOYDS & red/blu & 3200--10000 & 2.0 & 413 \\
2024-08-12 & 75.15 & IRTF+SpeX & ShortXD & 7000--25500 & 0.8 & 800* \\
2024-08-17 & 80.16 & LCO+FLOYDS & red/blu & 3200--10000 & 2.0 & 439 \\
2024-08-27 & 90.21 & LCO+FLOYDS & red/blu & 3200--10000 & 2.0 & 428 \\
2024-09-01 & 95.63 & Copernico 1.82m + AFOSC & VPH7 & 3400--8200 & 1.7 & 500* \\
2024-09-04 & 98.15 & LCO+FLOYDS & red/blu & 3200--10000 & 2.0 & 418 \\
2024-09-09 & 103.65 & CAHA + CAFOS 2.2 & red-400 & 3200--9000 & 1.0 & 300* \\
2024-09-12 & 106.16 & LCO+FLOYDS & red/blu & 3200--10000 & 2.0 & 393 \\
2024-09-19 & 113.12 & IRTF+SpeX & ShortXD & 7000--25500 & 0.8 & 800* \\
2024-09-20 & 113.86 & MMT + MMIRS & Grism & 9900--15000 & 1.0 & 1000* \\
2024-09-20 & 114.15 & LCO+FLOYDS & red/blu & 3200--10000 & 2.0 & 321 \\
2024-09-27 & 120.71 & TNG + LRS & LR-B & 3200-9000 & 1.0 & 800* \\
2024-09-28 & 121.99 & LCO+FLOYDS & red/blu & 3200--10000 & 2.0 & 420 \\
2024-10-03 & 127.57 & TNG + LRS & VHR-R & 6180--7830 & 1.0 & 1500* \\
2024-10-03 & 127.60 & TNG + LRS & LR-B & 3200-9000 & 1.5 & 800* \\
2024-10-06 & 130.03 & LCO+FLOYDS & red/blu & 3200--10000 & 2.0 & 398 \\
2024-10-10 & 134.58 & Copernico 1.82m + AFOSC & VPH6 & 3400--8200 & 1.7 & 500* \\
2024-10-24 & 148.07 & LCO+FLOYDS & red/blu & 3200--10000 & 2.0 & 423 \\
2024-10-31 & 155.57 & Copernico 1.82m + AFOSC & GR04 & 3400--8200 & 1.7 & 500* \\
2024-11-01 & 156.13 & LCO+FLOYDS & red/blu & 3200--10000 & 2.0 & 287 \\
2024-11-05 & 159.70 & MMT + BCH & 1200GPM & 5700-7000 & 1.0 & 3000* \\
2024-11-09 & 163.93 & LCO+FLOYDS & red/blu & 3200--10000 & 2.0 & 352 \\
2024-11-15 & 169.93 & IRTF+SpeX & Prism & 7000--25500 & 0.8 & 100* \\
2024-11-27 & 181.98 & LCO+FLOYDS & red/blu & 3200--10000 & 2.0 & 421 \\
2024-12-05 & 189.99 & LCO+FLOYDS & red/blu & 3200--10000 & 2.0 & 422 \\
2024-12-18 & 202.90 & LCO+FLOYDS & red/blu & 3200--10000 & 2.0 & 424 \\
2024-12-27 & 211.91 & LCO+FLOYDS & red/blu & 3200--10000 & 2.0 & 426 \\
2024-12-29 & 214.24 & HCT + HFOSC & Grism 7 & 3500--7700 & 2.0 & 616 \\
2025-01-02 & 217.91 & LCO+FLOYDS & red/blu & 3200--10000 & 2.0 & 424 \\
2025-01-10 & 225.91 & LCO+FLOYDS & red/blu & 3200--10000 & 2.0 & 440 \\
2025-01-19 & 235.51 & DOT + ADFOSC & 676R & 4200--8800 & 1.6 & 754 \\
2025-05-22 & 358.24 & LCO+FLOYDS & red/blu & 3200--10000 & 2.0 & 355 \\
2025-06-01 & 368.21 & LCO+FLOYDS & red/blu & 3200--10000 & 2.0 & 378 \\
2025-06-05 & 371.70 & MMT + BCH & 1200GPM & 5700-7000 & 1.0 & 3000* \\
2025-06-07 & 373.70 & SphereX & -- & 6000--30000 & -- & 100* \\
2025-06-22 & 389.23 & LCO+FLOYDS & red/blu & 3200--10000 & 2.0 & 391 \\
2025-11-27 & 546.70 & SphereX & -- & 6000--30000 & -- & 100* \\
\hline
\end{tabular}
\tablenotetext{\dagger}{The phase is given with respect to the explosion epoch of JD 2460459.8.}
\end{table*}

\label{lastpage}
\end{document}